\documentclass[12pt]{article}

\usepackage{fancybox} 
\usepackage{fancyhdr}
\usepackage{latexsym} 
\usepackage{amssymb}
\usepackage{graphicx}
\usepackage{epsf}

\newcommand{\nn}{\nonumber}
\def\sech{\mathop{\mathrm{sech}}\nolimits}
\def\be{\begin{eqnarray}}
\def\ee{\end{eqnarray}}

\def\q{\quad}

\def\dag{\dagger}

\def\st{\stackrel}

\def\lra{\longrightarrow}

\def\Ra{\Rightarrow}
\def\ra{\rightarrow}
\def\p{\partial}
\def\a{\alpha}

\def\m{\mu}
\def\n{\nu}

\def\th{\theta}

\def\l{\lambda}

\def\D{\mathcal{D}}

\def\z{\zeta}
\def\Re{{\rm Re}}
\def\Im{{\rm Im}}

\def\dpsi{\dot{\psi}}

\def\dr{\dot{r}}
\def\Lag{\mathcal{L}}

\def\e{{\rm e}}

\def\dth{\dot{\theta}}

\def\dR{\dot{R}}

\def\[{\bigl[}
\def\]{\bigl]}

\def\ve{\varepsilon}

\def\=:{=\hspace{-.7em}\raisebox{1.1ex}{.}\hspace{.1em}\raisebox{-0.2ex}{.} }

\def\S{\Sigma}
\def\tR{\tilde{R}}

\makeatletter
\@addtoreset{equation}{section}

\renewcommand{\thefootnote}{\fnsymbol{footnote}}
\makeatother

\begin{document}
\thispagestyle{empty}

\begin{flushright}
TIT/HEP--510 \\
{\tt hep-th/0310189} \\
October, 2003 \\
\end{flushright}
\vspace{3mm}

\begin{center}
{\Large
{\bf 
Exact Wall Solutions in $5$-Dimensional SUSY QED 
} 
\\
\vspace{2mm}{\bf 
at Finite Coupling 
} 
} 
\\[12mm]
\vspace{5mm}

\normalsize
  {\large \bf 
Youichi~Isozumi}
\footnote{\it  e-mail address: 
isozumi@th.phys.titech.ac.jp
}, 
  {\large \bf 
 Keisuke~Ohashi 
}\footnote{\it  e-mail address: 
keisuke@th.phys.titech.ac.jp
}, 
~and~~  {\large \bf 
Norisuke~Sakai}
\footnote{\it  e-mail address: 
nsakai@th.phys.titech.ac.jp
} 

\vskip 1.5em

{ \it Department of Physics, Tokyo Institute of 
Technology \\
Tokyo 152-8551, JAPAN  
 }
\vspace{15mm}
{\bf Abstract}\\[5mm]
{\parbox{13cm}{\hspace{5mm}
A series of exact BPS solutions are found for single and double 
domain walls in ${\cal N}=2$ supersymmetric (SUSY) QED 
for finite gauge coupling constants. 
Vector fields are found to be massive, although it is localized 
on the wall. 
Massless modes can be assembled into a chiral scalar multiplet of 
the preserved ${\cal N}=1$ SUSY, 
after an appropriate gauge choice. 
The low-energy effective Lagrangian for the massless 
fields is 
obtained for the finite gauge coupling. 
The inter-wall force is found to be much stronger 
than the known infinite coupling case. 
The previously proposed expansion in inverse powers of the 
gauge coupling has pathological oscillations, and 
does not converge to the correct finite 
coupling result. 
}}
\end{center}
\vfill
\newpage
\setcounter{page}{1}
\setcounter{footnote}{0}
\renewcommand{\thefootnote}{\arabic{footnote}}

\section{Introduction}\label{INTRO}


In recent years, much efforts have been devoted to the 
brane world scenario \cite{HoravaWitten}--\cite{RS}, 
where our world is supposed 
to be realized on a subspace such as a wall embedded in 
higher dimensional spacetime. 
On the other hand, supersymmetry (SUSY) has been one of 
the most fruitful ideas to build realistic unified models 
beyond the standard model \cite{DGSWR}. 
SUSY helps to obtain domain 
walls and other stable solutions by requiring 
 a part of SUSY to be preserved by the field configuration. 
Then the configuration has minimum energy with the given 
boundary condition and is 
automatically a solution of field equations. 
Wall 
configurations can preserve half of SUSY charges 
\cite{CQR}--\cite{EMSS}, which are called $\frac{1}{2}$ 
BPS states \cite{WittenOlive}. 
To obtain a wall with the four-dimensional world volume, 
we need to consider a fundamental theory in five or more 
spacetime dimensions. 
If such a theory are supersymmetric, 
it must have at least eight SUSY charges \cite{WB}.
Simplest theories with eight SUSY 
consist of hypermultiplets. 
The nontrivial interactions of hypermultiplets 
require either nonlinearity of kinetic term 
(nonlinear sigma model) or gauge interactions
 \cite{SierraTownsend}.
The target spaces of such nonlinear sigma models 
are hyper-K{\"a}hler (HK) manifolds~\cite{Zu, AF1} and 
can most conveniently be obtained by using 
vector multiplets as Lagrange multiplier fields~\cite{RT,LR}. 
By introducing the 
Fayet-Iliopoulos (FI) term for the $U(1)$ 
Lagrange multiplier vector multiplet and 
mass terms for hypermultiplets, 
one can obtain nontrivial potential terms with $N$ 
discrete SUSY vacua for $N$ hypermultiplets. 
Domain wall solutions have been obtained in these massive 
nonlinear sigma models with the target space 
$T^*{\bf C}P^{N-1}$ \cite{GTT1}--\cite{AFNS} and their 
generalizations \cite{EFNS}.
One of the most convenient methods to obtain 
these nonlinear 
sigma models is to take a limit of infinite gauge 
coupling from the usual minimal gauge interactions 
of hypermultiplets, namely massive ${\cal N}=2$ SUSY QED. 
However, it has been known that 
the massive ${\cal N}=2$ SUSY QED itself 
has discrete SUSY vacua 
and may allow domain walls even 
with finite gauge coupling 
\cite{Tong}--\cite{KakimotoSakai}. 

The massive nonlinear sigma models with multi-flavor 
have been found to 
give BPS multiple wall solutions containing 
a number of parameters, which 
represent the relative positions and phases of 
these multiple walls and are 
called moduli \cite{Tong}. 
When these moduli parameters are promoted to 
functions of the world volume coordinates of the 
walls, one can obtain the low-energy 
effective Lagrangian of moduli fields, 
which are massless \cite{Manton, RSVV, GTT}. 
It has been noted that gauge field should acquire nontrivial 
configuration when these moduli depend on the world volume 
coordinates \cite{ShifmanYung}. 
It has also been proposed that an expansion in inverse 
powers of gauge coupling may be useful to obtain 
the moduli dynamics at finite gauge coupling \cite{Tong}. 

To fulfill the goals of the brane-world scenario, 
we need to localize particles of the standard model to 
``branes'', such as domain walls. 
It has been a notoriously difficult problem to obtain 
massless localized gauge field on a wall 
\cite{LED, DvaliShifman2, MaruSakai}.
The general arguments \cite{LED, DvaliShifman2} and 
an explicit model \cite{MaruSakai} suggest that 
gauge fields may be localized on a wall, but will not 
be massless in the massive ${\cal N}=2$ SUSY QED 
with finite gauge coupling. 
Nevertheless, it is still interesting to study explicitly 
the role and the properties of dynamical gauge fields 
in the interacting system of the vector multiplet and 
hypermultiplets. 
In this respect, it should be useful to obtain exact 
solutions with finite gauge couplings 
\cite{KakimotoSakai}
\footnote{
In Ref.\cite{KakimotoSakai}, an exact solution of domain wall 
junction has been obtained for the massive ${\cal N}=2$ SUSY QED 
with the {\it finite gauge coupling} for the first time. 
Therefore it automatically implies an exact solution of domain wall 
for finite gauge coupling. 
}. 
In the four-dimensional massive ${\cal N}=2$ 
SUSY QED which can be obtained by a dimensional reduction of 
our model, it has been argued that a 
massless gauge field is obtained by dualizing the 
massless Nambu-Goldstone scalar \cite{ShifmanYung} 
\footnote{
They are interested in an ${\cal N}=1$ gauge theory 
obtained from ${\cal N}=2$ $SU(2)$ gauge theory deformed 
by an adjoint scalar mass term. 
However, they observed the model reduces to our ${\cal N}=2$ 
SUSY QED to the leading order of the mass deformation 
parameter. 
}. 
While this is an intriguing result, the gauge field as a dual of 
a compact scalar is an intrinsically three-dimensional peculiarity 
which is not valid to the more realistic situation such 
as our five-dimensional theory. 
In another paper, we show that massless localized gauge fields 
can be obtained by coupling our model to a tensor multiplet \cite{IOS}.

The purpose of our paper is to 
work out exact solutions of BPS single and multiple walls 
with finite gauge coupling in massive ${\cal N}=2$ SUSY QED 
in five dimensions and to study the fluctuations on the 
domain wall background.  
We obtain a series of exact solutions for single as well 
as multiple walls provided gauge coupling has particular 
values relative to the ratio of the mass parameters and 
FI parameters. 
We also show that the gauge multiplet is massive 
for any finite gauge coupling, 
irrespective of the possible existence of exact solutions. 
This is consistent with the previous general arguments 
\cite{LED, DvaliShifman2} 
and explicit examples\cite{MaruSakai}. 
In our model, the massless Nambu-Goldstone scalar still 
exist, but are apparently unrelated to the dual gauge field. 
Instead, we obtain the Nambu-Goldstone scalars and the 
Nambu-Goldstone fermions which form a 
chiral scalar multiplet of the remaining ${\cal N}=1$ 
SUSY, by choosing an appropriate gauge. 
Moreover we have only massive spectra for the vector field. 
These results are in contrast to the previous results 
of the massive 
${\cal N}=2$ SUSY $U(1)$ gauge theory in four dimensions, 
where the Nambu-Goldstone scalar is identified as the 
dual of the three-dimensional gauge field 
\cite{ShifmanYung}. 
As a result of exact solutions, we also obtain 
the low-energy effective Lagrangian for massless 
moduli fields for finite gauge couplings, 
by promoting 
the moduli parameters into four-dimensional massless fields 
\cite{Manton, RSVV, GTT}. 
In the process of carrying out the program, we observe that 
the gauge field should acquire nontrivial configuration 
with nonvanishing field strength around the wall, 
confirming the observation in Ref.\cite{ShifmanYung}. 
Moreover, we find that this nontrivial gauge field 
configuration should be determined by means of the field 
equation of gauge field. 
We compute also some lower order terms of 
the previously proposed 
expansion in inverse 
powers of gauge coupling \cite{Tong}, and 
compare the expansion with our exact result. 
We find that the expansion 
oscillates wildly and does not converge to the exact result 
in any smooth or uniform way. 
Therefore the expansion has a difficulty or at least subtlety 
in extracting physical quantities.


In Sect.~2, we introduce our model of massive ${\cal N}=2$ 
SUSY QED and obtain its SUSY vacua and BPS equations. 
In Sect.~3, we study the single wall in the case of two 
hypermultiplets ($N=2$), and work out fluctuation spectra to 
show that there are massless Nambu-Goldstone modes, 
but that the vector field has only massive fluctuations. 
In Sect.~4, a series of exact solutions are constructed 
for $N-1$ walls in the case of $N$ hypermultiplets 
and obtain the low-energy effective Lagrangian for 
the massless fields. 
We also work out a few lower order terms of the expansion in 
inverse powers of gauge coupling and compare it to 
our exact solution.

\section{Massive ${\cal N}=2$ SUSY QED and BPS equations}
\label{model}

The simplest building block of ${\cal N}=2$ SUSY theory 
in five dimensions 
is hypermultiplets which consist of  $SU(2)_R$ doublets of 
complex scalar fields $H^{iA}$, Dirac fields $\psi^A$ and 
complex auxiliary field $F_i^A$, where 
$i=1, 2$ stands for $SU(2)_R$  doublet indices and $A=1, \cdots, N$ 
stands for flavours. 
For simplicity, we assume that these $N$ hypermultiplets 
have the same $U(1)$ charge, say, unit charge. 
The $U(1)$ vector multiplet consists of 
 a gauge field $W_M$, a real scalar field $\S$, $SU(2)_R$ 
doublet of gauginos $\l^i$ and $SU(2)_R$ triplet of 
real auxiliary fields $Y^a$, where 
$M,N=0,1,\cdots,4$ denote space-time indices, and 
$a=1,2,3$ denotes $SU(2)_R$ triplet index, 
respectively.
In this work, we shall consider a model with the minimal kinetic 
term for hypermultiplets and vector multiplets. 
The ${\cal N}=2$ SUSY allows only a few parameters 
in our model : 
the gauge coupling $g$, the mass of the $A$-th 
hypermultiplet $m_A$, and the FI parameters $\zeta^a$. 
The FI parameters are real and transforms as a triplet 
under $SU(2)_R$. 
Then the 
bosonic part of our Lagrangian 
reads \cite{Tong}--\cite{KakimotoSakai}, \cite{Hebecker}
\be
\mathcal{L}_{\rm boson}
&\!\!\!=&\!\!\!
-\frac{1}{4g^2}(F_{MN}(W))^2+\frac{1}{2g^2}({\partial}_M\Sigma )^2 
+(\D_MH)^\dagger_{iA}(\D^MH^{iA})-H^\dagger_{iA}(\Sigma-m_A)^2H^{iA} 
\nn\\
&\!\!\!&\!\!\!
+\frac{1}{2g^2}(Y^a)^2 - \z^aY^a 
+ H^\dagger_{iA}(\sigma^aY^a)^i{}_jH^{jA} +F^{\dagger i}_AF_i^A ,
\label{5DW-3.1}
\ee
where a sum over repeated indices is understood, 
$F_{MN}(W)=\partial_M W_N -\partial_N W_M$, 
covariant derivative is defined as $\D_M =\p_M + iW_M$, 
and our metric is $\eta_{MN}=(+1, -1, \cdots, -1)$. 

To ensure the discreteness of SUSY vacua, 
we need to make all hypermultiplet masses nondegenerate. 
Without loss of generality, we assume the following 
ordering of the hypermultiplet mass parameters 
\be
m_{A+1}<m_A\label{5DW-3.6}
\ee
for all $A$. 
When all hypermultiplet masses are nondegenerate, 
the symmetry of our model reduces to 
$
U(1)^{N}
$. 
The diagonal $U(1)$ is gauged by the vector multiplet 
$W_M$, and the remaining $U(1)^{N-1}$ is the global 
symmetry. 

The easiest way to find SUSY vacua is to explore the condition of 
vanishing vacuum energy. 
To facilitate the procedure, 
let us first write down equations of motion of auxiliary 
fields $Y^a$ and $F_i^A$ 
\be
Y^a&\!\!\!=&\!\!\!
g^2[ \z^a-H^\dagger_{iA}(\sigma^a)^i{}_jH^{jA}], \\
F_i^A&\!\!\!=&\!\!\!0 .
\label{eq:EOMauxilary}
\ee
After eliminating the auxiliary fields, we obtain the 
potential $V_{\rm pot}$ 
\be
\mathcal{L}_{\rm boson}
&\!\!\!
=
&\!\!\!
-\frac{1}{4g^2}(F_{MN}(W))^2+\frac{1}{2g^2}({\partial}_M\Sigma )^2 
+(\D_MH)^\dagger_{iA}(\D^MH^{iA})-
V_{\rm pot},
\ee
\begin{equation}
V_{\rm pot} \equiv \frac{g^2}{2}
\left[ \z^a-H^\dagger_{iA}(\sigma^a)^i{}_jH^{jA}\right]^2
+ H^\dagger_{iA}(\Sigma-m_A)^2H^{iA}. 
\end{equation}
The vanishing vacuum energy is achieved by requiring 
\begin{equation}
H^\dagger_{iB}(\sigma^a)^i{}_jH^{jB}=\z^a, 
\label{eq:susyvac1}
\end{equation}
for each $a=1, 2, 3$ (the flavor index $B$ is summed) and 
\begin{equation}
(\Sigma-m_A)H^{iA}=0,
\label{5DW-3.10}
\end{equation}
for each flavor $A$ (the flavor index $A$ is not summed).
These SUSY vacuum conditions guarantee the full preservation of SUSY 
and can also be derived by requiring 
the SUSY transformation of fermions to vanish as we see 
immediately. 

By making an $SU(2)_R$ transformation, we can always bring 
the FI parameters to the third direction without loss of 
generality 
\be
\z^a=(0,0,\z), \qquad \zeta >0.
\ee
In this choice, we find $N$ discrete SUSY vacua ($A=1, \cdots, N$) 
explicitly 
by solving Eqs.(\ref{5DW-3.10}) and (\ref{eq:susyvac1}) as 
\be
\S = m_A,&& |H^{1A}|^2=\z,\q H^{2A}=0 , \nn\\
&& H^{1B}=0 ,\q\q H^{2B}=0, \q\q (B\not = A). 
\label{5DW-3.11}
\ee

Since fermions are assumed to vanish in the wall configuration, 
we need to examine only SUSY transformations of fermions to find a 
configuration preserving a part of SUSY. 
Gaugino $\lambda^i$ and hyperino $\psi^A$ transforms as\footnote{ 
Our gamma matrices are $4\times 4$ matrices and are defined as : 
$\{ \gamma^M,\gamma^N\}=2\eta^{MN }$, 
$\gamma^{MN}\equiv \frac{1}{2}[\gamma^M,\gamma^N]=\gamma^{[M}\gamma^{N]}$, 
$\gamma^5\equiv i\gamma^0\gamma^1\gamma^2\gamma^3=
-i \gamma^4$. 
}
\be
\delta_\varepsilon \lambda^i 
&\!\!\!=&\!\!\! \Bigl(\frac{1}{2}\gamma^{MN}F_{MN}(W)
+\gamma^M{\partial}_M\Sigma \Bigr)\varepsilon^i
+i\Bigl(Y^a\sigma^a\Bigr)^i{}_j\varepsilon^j, 
\label{eq:gauginoSUSY}
\\
 \delta_\varepsilon \psi^A 
&\!\!\!=&\!\!\! -i\sqrt{2} \Bigl[ \gamma^M\D_MH^{iA} 
+i(\Sigma-m_A) H^{iA} \Bigl] \epsilon_{ij}\varepsilon^j
+\sqrt{2}F_i^A\varepsilon^i .
\label{eq:SUSYtrans}
\ee
To obtain a  wall solution, 
we assume the configuration to depend only on the coordinate of 
one extra dimension, which we denote as $y \equiv x^4$. 
We also assume the 
four-dimensional Lorentz invariance in the world volume coordinates 
$x^\mu=(x^0, \cdots, x^3)$, which implies 
\be
F_{MN}(W)=0 . 
\label{5DW-3.5}
\ee
Since we are interested in the ${1 \over 2}$ BPS configuration, we 
require the above SUSY transformations (\ref{eq:SUSYtrans}) 
to vanish for half of 
the Grassman parameters specified by 
\be
P_+ \ve^1 =0, \q P_-\ve^2 =0 ,\label{5DW-3.13}
\ee
where $P_{\pm}\equiv (1 \pm \gamma_5)/2$ are the chiral 
projection operators. 
Finally we need to eliminate the auxiliary fields $Y^a$ and 
$F^A_i$ by their algebraic equations of motion 
(\ref{eq:EOMauxilary}) to make the BPS condition 
as first order differential 
equations for physical fields.  
Thus we obtain the ${1 \over 2}$ BPS equations for the massive 
${\cal N}=2$ SUSY QED as 
\be
\partial_y \S &\!\!\!=&\!\!\!g^2 
\left(\z-H_{1A}^\dagger H^{1A}+H_{2A}^\dagger H^{2A} \right),
\label{5DW-3.3}\\
2g^2 H_{2A}^\dagger H^{1A}
&\!\!\!=&\!\!\!
2g^2 H_{1A}^\dagger H^{2A}=0, 
\label{eq:BPS2}\\
\D_y H^{iA} &\!\!\!=&\!\!\!(m_A-\S )H^{iA}, 
\quad 
i=1, 2, 
\quad A=1, \cdots, N. 
\label{5DW-3.4}
\ee
One can easily see that the translation invariant vacuum requires 
the vanishing of the left-hand side of Eqs.(\ref{5DW-3.3}) and 
(\ref{5DW-3.4}), which implies the same 
condition as the full preservation of SUSY in Eqs.(\ref{5DW-3.10}) 
and (\ref{eq:susyvac1}). 

We are interested in the solution of these BPS equations which 
interpolate two different vacua in Eq.(\ref{5DW-3.11}). 
Since the BPS equation is a first order differential equation, 
the boundary condition dictates that $H^{2A}$ should 
vanish identically 
\be
H^{2A} (y) =0. 
\label{eq:H2A}
\ee

The energy density of the BPS solution can be found by making 
a Bogomolny completion of the energy functional 
\be
T_{\rm w} &\!\!\!=&\!\!\! 
\int^{\infty}_{-\infty}dy\Bigl\{ \frac{1}{2g^2}
\Bigl( \partial_y \S -g^2 (\z-|H^{1A}|^2 )\Bigl)^2 
+ |\D_y H^{1A} -(m_A-\S )H^{1A} |^2\Bigl\}\nn\\
&\!\!\!&\!\!\!
+\Bigl[ \z\S + (m_A-\S ) 
|H^{1A}|^2 \Bigl]^{y=+\infty }_{y=-\infty }. 
\ee
yielding 
the BPS wall tension 
for solutions interpolating the $A$-th vacuum and the $B$-th vacuum 
\be
T_{\rm w} =\z (m_A -m_B),
\label{eq:total-tension}
\ee
assuming $A > B$. 
The structure of the above BPS equations shows that 
only those hypermultiplet scalars $H^{1C}\ne 0$ with 
$A \ge C \ge B$ have nonvanishing values besides 
the vector multiplet scalar $\Sigma$ \cite{GTT}, 
\cite{EFNS}. 
Defining $l \equiv A-B$, we call such a BPS configuration 
as $l$ wall solutions, since it represents $l$ separate walls 
at least when these walls are sufficiently far apart, 
as we shall see in the Sect.\ref{sc:multi-wall}.

\section{Single wall BPS solutions}\label{sc:single-wall}

\subsection{Exact solutions for single wall}

Since single wall interpolates two adjacent SUSY vacua, 
we shall assume $N=2$ (two hypermultiplets), without loss 
of generality. 
For simplicity, we take mass parameters of the 
hypermultiplets 
as 
\be
m_1=-m_{2}\equiv m. 
\label{eq:single-wall-mass}
\ee
The boundary 
conditions for BPS equations are given by 
\be
\Sigma(-\infty)&\!\!\!
=&\!\!\!-m,\quad \Sigma(\infty)=m, 
\label{eq:vector-bound-cond}
\\
H^{11}(-\infty) &\!\!\!
=&\!\!\! 0,\quad H^{11}(\infty)=\sqrt{\z}, \nn\\
H^{12}(-\infty)&\!\!\!
=&\!\!\!\sqrt{\z},\quad H^{12}(\infty)=0 .
\label{eq:hyper-bound-cond}
\ee

The set of the BPS equations 
(\ref{5DW-3.3})--(\ref{5DW-3.4}) with 
the above boundary conditions 
(\ref{eq:vector-bound-cond})--(\ref{eq:hyper-bound-cond}) 
are known to be solved exactly 
for infinite gauge coupling  \cite{GTT}, 
\cite{EFNS}.  
Recently another exact solution has been found 
for finite gauge coupling, 
provided the gauge coupling $g$ 
satisfies the following relation with the ratio of the FI 
parameter $\zeta$ and the mass difference squared 
$(2m)^2$ \cite{KakimotoSakai}
\be
g^2 \z = 2m^2.
\ee
By generalizing this exact solution for finite gauge coupling, 
we find that there are exact solutions for the following 
values of gauge coupling $g$ 
\be
g^2\z \equiv \frac{8m^2}{k^2}, 
\label{eq:solvable}
\ee
with appropriate integers $k$. 
We shall denote the single wall 
solution for $k$ as $S_{k}(m)$. 
The integer $k$ 
indicates the value of the gauge coupling $g$ through 
the relation (\ref{eq:solvable}), and 
the mass parameter $m$ 
indicates the tension $T_{\rm w}$ of the single wall in unit of 
the FI parameter as $T_{\rm w}=2\zeta m$. 
The infinite gauge coupling corresponds to the case $k=0$, and 
the previously obtained finite gauge coupling case to $k=2$ 
\cite{KakimotoSakai}. 
We have explicitly obtained exact solutions in the case of 
$k=0, 2, 3, 4$. 

\begin{figure}[htb]
\begin{center}
\leavevmode
\epsfxsize=6cm
\epsfysize=4cm
\begin{picture}(100,100)(50,10)
\hspace{-10ex}
\epsfbox{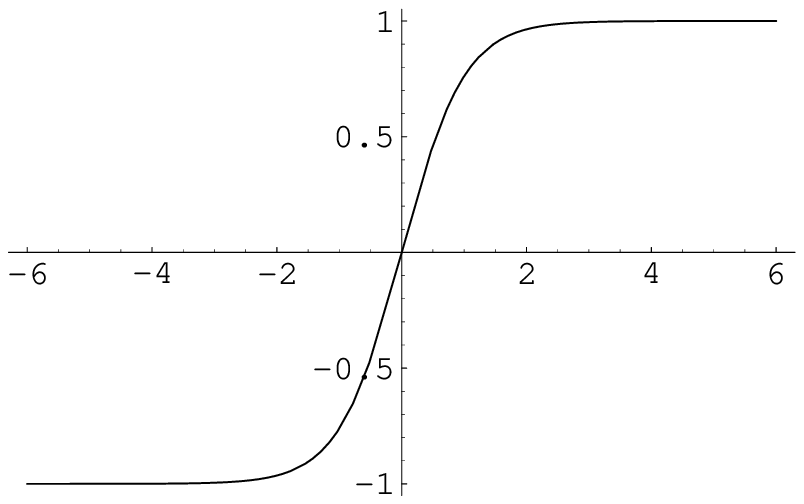}
\put(-10,60){$my$}
\put(-90,120){$\S/m$}
\put(-90,-20){a)}
\end{picture}
\epsfxsize=6cm
\epsfysize=4cm
\label{fig1-1}
\begin{picture}(100,100)(-10,10)
\epsfbox{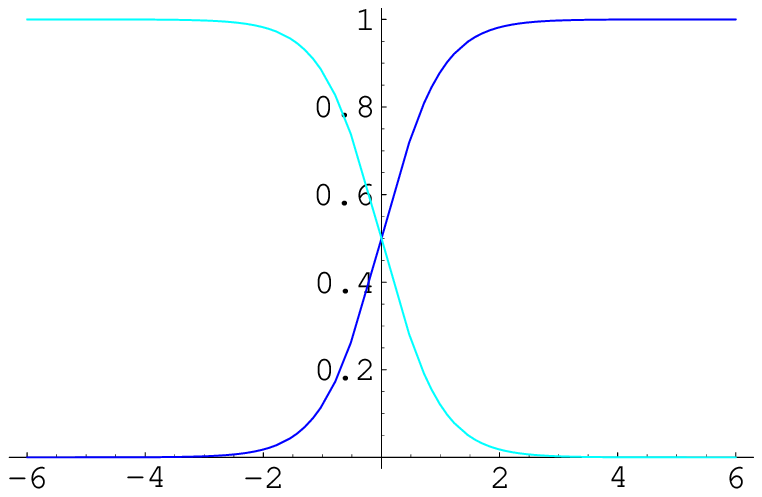}
\put(-10,15){$my$}
\put(-160,90){$H^{12}$}
\put(-25,90){$H^{11}$}
\put(-100,120){$\sqrt{2}gH^{1A}/m$}
\put(-90,-20){b)}
\end{picture}
\end{center}
\vspace{0.5cm}
\caption{The exact BPS wall solution $S_2(m)$ 
for finite gauge 
coupling $g={\sqrt2 m}/{\sqrt{\zeta}}$ and the tension 
$T_{\rm w}=2m\zeta$. 
a) Scalar field $\Sigma(y)$ of vector multiplet in 
Eq.(\ref{eq:single-wall-sigma}) divided by the 
mass parameter $m$ 
as a function of the 
coordinate $y$ times $m$. 
b) Hypermultiplet scalar field $H^{11}(y)$ 
in Eq.(\ref{eq:single-wall-H1}) as a function of the 
coordinate $y$ times $m$ 
(solid line),  
and $H^{12}(y)$ in Eq.(\ref{eq:single-wall-H2}) 
 as a function of the 
coordinate $y$ times $m$ (dotted line). 
}
\label{fig:BPS-single-wall}
\end{figure}

In this section, we will describe the exact solution $S_{2}(m)$ 
for $k=2$ as the simplest 
example, which plays an 
important role in our paper on massless localized 
vector field \cite{IOS} : 
\be
\Sigma(y)&\!\!\!=&\!\!\!m \tanh my, 
\label{eq:single-wall-sigma}\\
H^{11}(y)
&\!\!\!=&\!\!\!\frac{m}{\sqrt{2}g} e^{+ my} \sech my,
\label{eq:single-wall-H1}\\
H^{12}(y)
&\!\!\!=&\!\!\!\frac{m}{\sqrt{2}g} e^{- my} \sech my.
\label{eq:single-wall-H2}
\ee
Let us recall that we can introduce two arbitrary integration 
constants in these solutions. 
We can have $y \rightarrow y-y_0$ corresponding to the spontaneously 
broken translation invariance, 
and we can also have the multiplication by 
a phase ${\rm e}^{-i\alpha}$ ( ${\rm e}^{i\alpha}$) 
for hypermultiplets $H^{11}$ ($H^{12}$) corresponding to 
the spontaneously broken global $U(1)$ invariance. 
They constitute two moduli of the solution.  
The $y_0$ has a physical meaning of the position of the wall. 
The vector multiplet scalar $\Sigma$ and the hypermultiplet 
scalars $H^{11}$ and $H^{12}$ of the solution $S_2 (m)$ 
are illustrated as a function of the coordinate $y$ in the 
extra dimension in Fig.\ref{fig:BPS-single-wall}.
We will present other exact solutions of single wall 
after introducing a slightly different notation suitable also for 
multiple walls.

\vspace{1cm}
\vspace{1cm}

\subsection{Fluctuation around the BPS single wall background}
\label{sc:fluctuation}

Let us now turn our attention to spectra of fluctuations around the 
BPS single wall background. 
These fluctuation fields can be expanded into modes 
defined on the background. 
Among various modes, we are particularly interested in two 
kinds of modes : 
massless modes (zero modes) and fluctuation of vector fields. 
The former modes play essential role in discussing dynamics 
on the background at low-energies. 
The latter is to examine the role played by the dynamical 
vector field at finite coupling, rather than 
the role as the Lagrange 
multiplier field in the case of corresponding nonlinear sigma 
models.

Zero modes usually arise from several reasons. 
The first one is the Nambu-Goldstone modes as a result of 
the spontaneous breaking of continuous global symmetry. 
The second possibility is the result of remaining symmetry, 
such as the conserved SUSY \cite{HNOO}. 
The third possibility is the result of index theorems, such 
as fermion zero modes around instantons and other solitons 
\cite{JackiwRebbi}. 

To illustrate the method, we shall work on the fluctuations 
on the background of the simplest exact solution $S_{2}(m)$ 
in Eqs.(\ref{eq:single-wall-sigma})--(\ref{eq:single-wall-H2}). 
In the case of a single wall, such as $S_{2}(m)$, 
the only spontaneously broken global internal symmetry 
is the $U(1)$ symmetry which rotates the phase of 
hypermultiplets 
oppositely 
\be
H^{i1}\rightarrow 
{\rm e}^{-i\alpha}
H^{i1}, \quad 
H^{i2}\rightarrow 
{\rm e}^{i\alpha}
H^{i2} . 
\label{eq:globalU1}
\ee
There are two bosonic global symmetry which are 
spontaneously broken : 
translation and the global $U(1)$ in Eq.(\ref{eq:globalU1}). 
The mode function of the Nambu-Goldstone boson 
corresponding to the spontaneously broken translation 
is given by 
differentiating 
the background field configuration with respect to the 
coordinate in extra dimension $y$. 
We shall denote the corresponding four-dimensional 
effective field 
as Re$\phi_0(x)$. 
Similarly the Nambu-Goldstone boson associated with 
the global $U(1)$ symmetry is given by acting an 
infinitesimal $U(1)$ transformation. 
We denote the corresponding four-dimensional 
effective field 
as Im$\phi_0(x)$. 
By writing out only zero modes, we thus obtain 
scalar fields of vector multiplet $\Sigma(x,y)$ 
and that of hypermultiplets $H^{1A}(x, y), A=1, 2$ 
\be
\Sigma(x, y) &\!\!\!
=&\!\!\! \langle \Sigma (y)\rangle -
\partial_y\langle \Sigma(y) \rangle {\rm Re}\phi_0(x),
\label{eq:NGbosonSigma}
\\
H^{11}(x, y)&\!\!\!=&\!\!\!\langle H^{11}(y)\rangle
-\partial_y \langle H^{11}(y)\rangle {\rm Re}\phi_0(x)
-im\langle H^{11}(y)\rangle  {\rm Im}\phi_0(x),
\label{eq:NGbosonH1}
\\
H^{12}(x, y)&\!\!\!=&\!\!\!\langle H^{12}(y)\rangle
-\partial_y \langle H^{12}(y)\rangle {\rm Re}\phi_0(x)
+im\langle H^{12}(y)\rangle {\rm Im}\phi_0(x),
\label{eq:NGbosonH2}
\ee
where we multiplied the factor $m$ in front of the 
$U(1)$ Nambu-Goldstone mode wave function for 
later convenience, and 
the $\langle \cdots \rangle$ denotes the background 
field configuration. 
These configurations together with vanishing 
gauge field $W_M=0$ 
are solutions of the BPS equation 
if Re$\phi_0$ 
and Im$\phi_0$ are constants (moduli) and hence 
automatically satisfy the field equations. 
However, we are now interested in the case where 
these  Re$\phi_0$ 
and Im$\phi_0$ are four-dimensional fields, 
which depend on the coordinate 
$x^\mu$ of the world volume. 
Then these field configurations are no longer 
BPS configurations, nor satisfy  the field equations. 
Since four-dimensional effective fields are 
defined as solutions of the field equations, we should 
demand these field configurations to satisfy 
the linearized field equations. 
Since $W_M$ must vanish for constant $\phi_0$, 
it should be proportional to $\partial_\mu \phi_0(x)$. 
Therefore the Lorentz invariance 
in the four-dimensional 
world volume dictates 
\begin{equation}
W_4(x, y)=0, 
\label{eq:W4-sol}
\end{equation}
and 
\begin{equation}
W_\mu(x, y)=w(y)\partial_\mu {\rm Im}\phi_0(x). 
\label{eq:Wmu-sol1}
\end{equation}
with some function $w(y)$ 
to the linearized order which we are interested in. 
Then the linearized field equations for the vector multiplet 
scalar $\Sigma$, hypermultiplet scalars $H^{1A}$ 
are satisfied by 
\begin{equation}
\partial_\mu \partial^\mu \phi_0(x)=0, 
\label{eq:phi-sol}
\end{equation}
which just means that these Nambu-Goldstone fields 
are massless. 

The remaining linearized field equation for the gauge 
field $W_M$ reads 
\be
\p_\n\p^\n W_\m -\partial_y^2 W_\m
-\p_\m (\p_\n W^\n )+\p_\m \p_yW_4 
=-2g^2 
\left(\langle H^{1A}\rangle \p_\m \Im h^{1A} 
+|\langle H^{1A}\rangle |^2W_\m   \right), 
\label{eq:Wmu-EOM}
\ee
\be
\p_\n\p^\n W_4 
-\partial_y (\p_\n W^\n )
=2g^2 
\left(\partial_y \langle H^{1A}\rangle \Im h^{1A} 
-\langle H^{1A}\rangle \p_y \Im h^{1A} 
-|\langle H^{1A}\rangle |^2W_4   \right). 
\label{eq:W4-EOM}
\ee
The solutions for $W_M$ in Eqs.(\ref{eq:W4-sol}) and 
(\ref{eq:Wmu-sol1}) are enough to satisfy 
the field equation for $W_4$ in Eq.(\ref{eq:W4-EOM}). 
The remaining field equation (\ref{eq:Wmu-EOM}) 
for $W_\mu$ determines the $y$ dependence $w(y)$ 
of the gauge field $W_\mu(x, y)$ in 
Eq.(\ref{eq:Wmu-sol1}) as 
\begin{equation}
W_\mu(x, y)=\langle \Sigma(y) \rangle 
\partial_\mu {\rm Im}\phi_0(x). 
\label{eq:Wmu-sol2}
\end{equation}
This solution shows that the corresponding field strength 
is nonvanishing around the wall 
\begin{equation}
F_{\mu 4}(W)(x, y)=-\partial_y\langle \Sigma(y) \rangle 
\partial_\mu {\rm Im}\phi_0(x). 
\label{eq:Fmuy}
\end{equation}
In a similar model in four-dimensions, 
it has been observed that the gauge field must be 
nontrivial and field strength must be nonzero 
around the wall if the moduli depends on world 
volume coordinates \cite{ShifmanYung, Tong}. 
Our result in five dimensions are in agreement 
with their observation. 
Our new point is perhaps that the necessary gauge field 
configuration can be determined explicitly 
by using the linearized field equations for gauge fields.

SUSY is another global symmetry which is spontaneously broken. 
We can obtain the Nambu-Goldstone fermion corresponding to 
the broken SUSY by evaluating the SUSY transformations 
of fermions on the background. 
One set of $SU(2)$-Majorana spinor $\lambda^i, i=1, 2$ 
in five dimensions can be 
decomposed into two Majorana spinors $\lambda_+, \lambda_-$ 
in four dimensions as 
\be
\lambda_+=P_+\lambda^1+P_-\lambda^2, 
\quad 
\lambda_-=-P_-\lambda^1+P_+\lambda^2. 
\label{5dSYM-6130}
\ee
The SUSY transformations in Eqs.(\ref{eq:gauginoSUSY}) and 
(\ref{eq:SUSYtrans}) can be evaluated 
by using the BPS equations (\ref{5DW-3.3})--(\ref{5DW-3.4}) 
to yield for the gauginos 
\be
\delta_\varepsilon \lambda_+
&\!\!\!=&\!\!\!
 i\left[\partial_y \langle \Sigma \rangle
+ g^2\left(\zeta-|\langle H^{1A}\rangle |^2\right)\right]
\left(P_{+}\varepsilon_+ - P_{-}\varepsilon_+\right) 
=2i\partial_y \langle\Sigma\rangle 
\left(P_{+}\varepsilon_+ - P_{-}\varepsilon_+\right) 
\nn\\
\delta_\varepsilon \lambda_-
&\!\!\!=&\!\!\!
 i\left[\partial_y \langle\Sigma\rangle 
- g^2\left(\zeta-|\langle H^{1A}\rangle|^2\right)\right]
\left(P_{-}\varepsilon_- + P_{+}\varepsilon_-\right) 
=0
. 
\label{eq:gaugino-SUSY}
\ee
Since $\varepsilon_-$ is preserved, it drops out from these 
transformations. 
If we evaluate explicitly for our exact solution $S_{2}(m)$ in 
Eqs.(\ref{eq:single-wall-sigma})--(\ref{eq:single-wall-H2}), 
we obtain 
\be
\delta_\varepsilon \lambda_+
={2im^2 \over \cosh^2 my} \left(P_{+}\varepsilon_+ 
- P_{-}\varepsilon_+\right) , 
\quad 
\delta_\varepsilon \lambda_-
=0.
\ee
Similarly the hyperino 
component of the Nambu-Goldstone fermion wave function 
is  given by 
\be
\delta_\varepsilon \psi^A
&\!\!\!=&\!\!\!
\sqrt{2}\left[\partial_y \langle H^{1A}\rangle 
+ \left( \langle \Sigma\rangle -m_A\right)
\langle H^{1A}\rangle \right]P_+\varepsilon_- 
+
\sqrt{2}\left[-\partial_y \langle H^{1A}\rangle
+ \left(\langle\Sigma\rangle -m_A\right)
\langle H^{1A}\rangle\right]P_-\varepsilon_+ 
\nn\\
&\!\!\!=&\!\!\!
-2\sqrt{2}\partial_y \langle H^{1A}\rangle
P_-\varepsilon_+ 
.
\label{eq:hyperino-SUSY}
\ee
For the exact solution  $S_{2}(m)$ in 
Eqs.(\ref{eq:single-wall-sigma})--(\ref{eq:single-wall-H2}), 
it becomes explicitly as 
\be
\delta_\varepsilon \psi^1=-
\delta_\varepsilon \psi^2=-
{2m^2 \over g} {1 \over \cosh^2 my}P_-\varepsilon_+
.
\ee
The above transformation laws (\ref{eq:gaugino-SUSY}) 
and (\ref{eq:hyperino-SUSY}) show that 
$-4\varepsilon_+$ is proportional 
to the zero momentum component of the Nambu-Goldstone 
fermion. 
Therefore we can define the Nambu-Goldstone fermion field 
$\chi_0(x)$ of the 
four-dimensional effective low-energy theory on the 
world volume of the wall as 
\be
\lambda_+ (x, y) 
&\!\!\!=&\!\!\! 
-{i \over 2}\partial_y
\langle \Sigma(y)\rangle \gamma_5 \chi_0(x), 
\quad
\lambda_- (x, y)= 0, 
\label{eq:NGfermion_lambda}
\\
\psi^A(x, y)&\!\!\!=&\!\!\!
{1 \over \sqrt{2}}\partial_y
\langle H^{1A}(y)\rangle P_- \chi_0(x), 
\label{eq:NGfermion_psi}
\ee
suppressing to write massive modes. 
We see explicitly that the Nambu-Goldstone fermion appears 
only with the left-handed chirality in 
the hypermultiplets as is usually dictated by index theorems 
for fermions localized on domain walls \cite{JackiwRebbi}. 
We can also verify that 
the Nambu-Goldstone fermion satisfies the linearized 
equations of motion with a vanishing 
mass eigenvalue, 
by using the BPS equations (\ref{5DW-3.3})--(\ref{5DW-3.4}). 

Now let us consider the requirement of the 
preserved symmetry, especially SUSY. 
The SUSY transformation property under the preserved SUSY 
 $\varepsilon_-$ specified by Eq.(\ref{5DW-3.13}) is given by 
\be
\delta_{\varepsilon_-} W_\mu
&\!\!\!=&\!\!\!-i\bar{\varepsilon}_-\gamma_\mu\lambda_-,
\\
\delta_{\varepsilon_-} W_4
&\!\!\!=&\!\!\!\bar{\varepsilon}_-\lambda_+, 
\\
\delta_{\varepsilon_-} \Sigma
&\!\!\!=&\!\!\!-i\bar{\varepsilon}_-\gamma_5\lambda_+,
\\
\delta_{\varepsilon_-} H^{1A}
&\!\!\!=&\!\!\!-\sqrt{2}\bar{\varepsilon}_-P_-\psi^A,
\\
\delta_{\varepsilon_-} H^{2A}
&\!\!\!=&\!\!\!
-\sqrt{2}\bar{\varepsilon}_-P_+\psi^A\, . \label{5dSYM-726}
\ee
This transformation property does not fit well with the 
above massless particles of Nambu-Goldstone modes 
in Eqs(\ref{eq:NGbosonSigma})--(\ref{eq:Wmu-sol1}), 
(\ref{eq:NGfermion_lambda}), and (\ref{eq:NGfermion_psi}), 
especially for the vector multiplet. 
However, we should remember that we have a freedom to make 
gauge transformations to make the above massless particles 
in conformity with the SUSY transformation properties. 
Let us perform the following gauge transformation, 
which is proportional to the fluctuation field Im$\phi_0$ 
\be
W_4 &\!\!\!\longrightarrow&\!\!\! 
W_4-\partial_y \left( \langle \Sigma\rangle 
{\rm Im}\phi_0\right),\nn\\
W_\mu &\!\!\!\longrightarrow&\!\!\! 
W_\mu-\partial_\mu 
\left(\langle\Sigma\rangle {\rm Im}\phi_0\right),\nn\\
H^{1A}&\!\!\!\longrightarrow&\!\!\!
 \langle H^{1A}\rangle+i  \left( \langle \Sigma \rangle 
{\rm Im}\phi_0\right)\langle H^{1A}\rangle .
\ee
Then the hypermultiplet scalars become 
\be
H^{11}(x, y)&\!\!\!=&\!\!\!
\langle H^{11}( y)\rangle -\partial_y \langle H^{11}(y)\rangle
{\rm Re}\phi_0(x)
-i \left({m}-\langle \Sigma(y)\rangle \right) \langle H^{11}(y)\rangle 
{\rm Im}\phi_0(x)
\nn\\
&\!\!\!=&\!\!\!
\langle H^{11}(y)\rangle 
-\partial_y \langle H^{11}(y)\rangle 
\left({\rm Re}\phi_0(x)+ i {\rm Im}\phi_0(x)\right) ,
\label{eq:hyper1_zm_2}
\\
H^{12}(x, y)&\!\!\!=&\!\!\!\langle H^{12}(y) \rangle 
-\partial_y \langle H^{12}(y)\rangle
{\rm Re}\phi_0(x)
-i \left(-{m}-\langle \Sigma(y)\rangle \right) 
\langle H^{12}(y) \rangle
{\rm Im}\phi_0(x)
\nn\\
&\!\!\!=&\!\!\!
\langle H^{12}(y)\rangle 
-\partial_y \langle H^{12}( y)\rangle \left(
{\rm Re}\phi_0(x)
+i {\rm Im}\phi_0(x)\right). 
\label{eq:hyper2_zm_2}
\ee
Therefore we find that the two Nambu-Goldstone bosons corresponding 
to translation and $U(1)$ global rotation forms a complex scalar 
 $\phi_0(x)$ of a chiral scalar multiplet of the preserved four SUSY 
$\epsilon_-$. 
At the same time, the four-dimensional gauge fields $W_\mu$ vanishes, 
and vector multiplet scalar $\Sigma$ and the fourth (extra dimension) 
component of vector field $W_4$ fit into the complex scalar $\phi_0(x)$ 
\be
\Sigma(x, y) + i W_4(x, y) &\!\!\!
=&\!\!\! \langle \Sigma( y)\rangle 
-\partial_y \langle \Sigma( y)\rangle
\left({\rm Re}\phi_0(x)
+i{\rm Im}\phi_0(x)\right),
\label{eq:vector_NG1}
\\
W_\mu(x, y) &\!\!\!=&\!\!\! 0 .
\label{eq:vector_NG2}
\ee
These mode expansions together with the fermionic ones 
in Eqs.(\ref{eq:NGfermion_lambda}) and (\ref{eq:NGfermion_psi}) 
are now consistent with the preserved 
SUSY which defines the four-dimensional ${\cal N}=1$ SUSY 
transformations for the massless fields in the effective field 
theory 
\be
\delta_{\varepsilon_-} \phi_0(x)
=
\bar \varepsilon_- P_- \chi_0(x). 
\ee

Thus we find that all the Nambu-Goldstone bosons and 
fermions together form a chiral scalar multiplet 
$(\phi_0(x), \chi_0(x) )$ of the 
preserved SUSY. 
Moreover we do not find any index theorem to force us 
additional zero modes. 
Therefore we do not expect any more zero modes. 

\begin{figure}[ht]
\begin{center}
\leavevmode
\scalebox{0.8}{\includegraphics{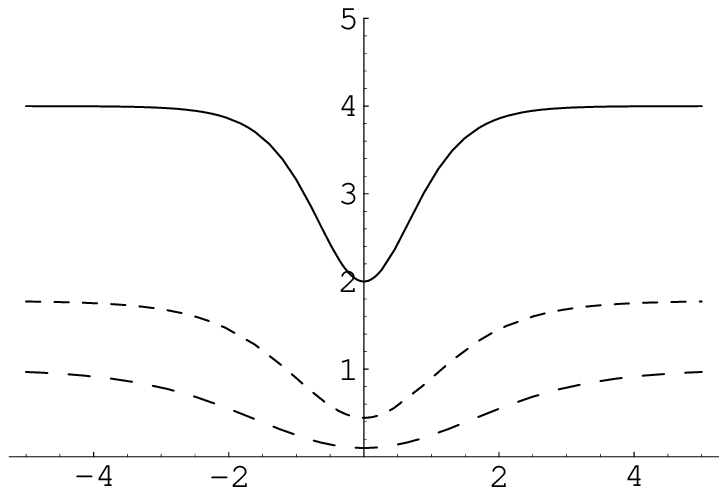}}
\begin{picture}(100,100)(0,0)
\put(0,5){$x$}
\put(-110,120){${V}(x)$}
\put(-10,95){$S_2(m)$}
\put(-10,45){$S_3(m)$}
\put(-10,25){$S_{4}(m)$}
\end{picture}
\vspace{0.5cm}
\caption{
The potential $V(x)$ of Eq.(\ref{5DW-1.20}) 
for $S_2(m)$ (solid line), $S_3(m)$ (dotted line), and 
$S_{4}(m)$ (dashed line) 
in the eigenvalue equation for the vector fluctuation 
(\ref{5DW-3.16}) as a function of extra dimension 
coordinate $y$.  
The mass parameter is taken to be $m=1$. 
 }
\label{fig2}
\end{center}
\end{figure}

Let us now turn to the issue of mass spectra, especially 
for vector fields. 
The linearized equation of motion for vector field 
$W_\mu$ and $W_4$ are given in Eqs.(\ref{eq:Wmu-EOM}) 
and (\ref{eq:W4-EOM}). 
By exploiting the $U(1)$ gauge invariance, we can 
always choose a gauge where $W_4(x, y)=0$ identically. 
We can also decompose vector field into 
the transverse part $\tilde W_\mu$ and longitudinal part 
$w$ 
\begin{equation}
W_\mu(x, y)=\tilde W_\mu(x, y) + \partial_\mu w(x, y), 
\quad \partial^\mu \tilde W_\mu=0. 
\label{eq:transvers-long}
\end{equation}
The transverse part of the vector field\footnote{
This component corresponds to the polazation states 
proportional to 
the polazation vector 
$\epsilon^i_\mu, i=1, 2,3$ defined by 
$\epsilon_\mu^i p^\mu=0$ in momentum space, 
representing a genuine four-dimensional 
vector field. 
}
is decoupled from the 
rest of the fluctuations because of the four-dimensional 
Lorentz invarince and gives the following mode 
decomposison 
with the mode function $a_n(y)$ for the mass 
eigenvalue $m_n$ 
of the four-dimensional effective field $W_\mu^{(n)}(x)$ 
\be
\tilde W_\mu(x, y)=\sum_n a_n(y) W^{(n)}_\mu(x),
\ee 
\be
-\p_y^2 a_n(y)+V(y)a_n(y)=m_n^2 a_n(y). \label{5DW-3.16}
\ee
where the potential $V(y)$ is defined by 
\be
V(y)=2g^2 |\langle H^{1A}\rangle |^2, 
\label{5DW-1.20}
\ee
and is illustrated in Fig.\ref{fig2} 
for the background $S_{2}(m)$ 
in Eqs.(\ref{eq:single-wall-sigma})--
(\ref{eq:single-wall-H2}), and $S_{3}(m)$ 
and $S_{4}(m)$, whose exact solutions are given 
explicitly in Sect.\ref{sc:exact-multi-wall}. 
\vspace{1cm}
These potentials have a dip near the wall 
and approaches $2g^2 \zeta$ at $y\rightarrow \pm \infty$. 
Therefore all the vector fluctuations become infinitely 
heavy as $g \rightarrow \infty$. 
Potentials are always positive definite, 
but have attractive 
forces due to a dip around the wall for any finite 
gauge coupling. 
It is likely that there may be discrete low mass 
states before reaching a continuum starting 
from $m^2=2g^2 \zeta$. 
The potential becomes infinite 
for the infinite gauge coupling. 
It is clear that the eigenvalue can never vanish : 
$m_n >0$. 
It is also easy to see that $m_n$ are of the order 
of $m$ for, 
say, $S_2 (m)$. 
This result is consistent with the general argument that the 
gauge field will obtain a mass of the order of the inverse 
width of the wall. 
Although the low-lying vector modes have presumably 
discrete spectra 
and are localized  around the 
wall, all these modes are massive. 
To obtain a massless vector field, we can add tensor multiplet\cite{IOS}. 

To find out the remaining field equations, we define 
fluctuations of field as 
$\Sigma =\langle \Sigma \rangle +s$, 
$\, H^{Ai}=\langle H^{Ai}\rangle +h^{Ai}$. 
The linearized field equations are given by 
\begin{equation}
0= 
\partial _M\partial ^Ms+2g^2\langle |H^{Ai}|^2\rangle s
+2g^2h^\dagger _{Ai}\langle (\Sigma -m_A)H^{Ai}\rangle 
+2g^2\langle H_{iA}^\dagger (\Sigma -m_A)\rangle h^{Ai}, 
\end{equation}
\begin{eqnarray}
0&\!\!\!=&\!\!\!
\partial _M\partial ^Mh^{Ai}+i(\partial ^MW_M)
\langle H^{Ai}\rangle +2iW_M\partial ^M
\langle H^{Ai}\rangle 
-g^2(\zeta -\langle |H^{Bj}|^2\rangle )h^{Ai}
\nonumber \\
&\!\!\!&\!\!\!
{}+g^2\langle H^{Ai}\rangle 
\left(\langle H_{Bj}^\dagger \rangle 
h^{Bj}+\langle H^{jB}\rangle h_{Bj}^\dagger \right)
+2(\langle \Sigma \rangle-m_A) \langle H^{Ai}\rangle s
+(\langle \Sigma \rangle -m_A)^2h^{Ai}. 
\label{eq:EOM-s-h}
\end{eqnarray}
These linearized field equations 
can be decomposed into three sets of coupled equations. 
The first set consists of the fluctuations of 
vector multiplet scalar 
$\Sigma$, and those of the 
real part of the hypermultiplet scalars 
Re$H^{1A}, \; A=1, 2$. 
The second set consists of imaginary part of 
 the hypermultiplet scalars 
Im$H^{1A}, \; A=1, 2$ and the longitudinal part of 
the vector field $w$ in Eq.(\ref{eq:transvers-long}). 
The third set consists of the lower components of 
the hypermultiplet scalars 
$H^{2A}, \; A=1, 2$. 
Among various massive modes, we find the following 
tower of massive modes as a solution of the first set of 
coupled equations 
\begin{eqnarray}
 s=\sum_n(\partial _ya_n(y))\phi _n(x), 
 \quad h^{A1}=
 -\sum_n\langle H^{A1}(y)\rangle a_n(y)\phi _n(x), 
\end{eqnarray}
with $h^{A2}=W_M=0$. 
This effective field $\phi_n(x)$ has the identical mass 
squared $m_n^2$ as the effective field $W_\mu^{(n)}(x)$ 
of the transverse vector in Eq.(\ref{5DW-3.16}). 
Moreover the mode function is also given by 
the same wave function 
$a_n(y)$ as the transverse vector. 
Therefore it is likely that these effective fields are 
related by the preserved symmetry, in particular 
the ${\cal N}=1$ SUSY. 
In fact, a massive vector multiplet of ${\cal N}=1$ SUSY 
should contain a scalar particle.

\section{BPS solution for Multiple walls}
\label{sc:multi-wall}

\subsection{Exact solutions with finite gauge couplings}
\label{sc:exact-multi-wall}

It is convenient to introduce a complex function $\psi(y)$ 
to solve the BPS equation for hypermultiplets 
(\ref{5DW-3.4}) 
\cite{Tong} 
\be
H^{1A}(y) 
&\!\!\!=&\!\!\! \sqrt{\z} 
\exp \Bigl( -\psi(y) +m_A (y-y_0)+\sum_{a=1}^{N-2}\a_A{}^a r_a \Bigl), 
\label{tong-13}
\ee
where the complex parameters $r_a, a=1, \cdots, N-2$ 
are collective coordinates 
arising as integration 
constants. 
Since two complex parameters among the integration constants 
can be absorbed by a shift of $\psi$ and $y_0$, we can choose 
the $N \times (N-2)$ fixed real matrix 
$\a_A{}^a$ to be of rank $N-2$. 
The BPS equation for hypermultiplets is equivalent 
to the following 
equation for $\psi(y)$ 
\be
\p_y \psi = \S + i W_4 . 
\label{tong-15}
\ee
Because of the vanishing field strength (\ref{5DW-3.5}), 
the vector field $W_4={\rm Im}\partial_y\psi$ 
is a pure gauge. 
However, when we consider dynamics of domain walls, 
this term will play an important role.\\
We have the following moduli parameters in the solution 
$(\ref{tong-15})$,
\be
y_0&\!\!\!=&\!\!\!Y_0
+i\th_0 ,
\\
r_a&\!\!\!=&\!\!\!R_a+i\th_a , 
\label{tong-16}
\ee
with one and $N-2$ complex dimensions, respectively. 
The $Y_0$ and $R_a$ are related to the center of mass 
and relative positions of the $a$-th domain wall, respectively, and 
$\theta_0$ and $\theta_a$ to the overall phase and 
relative phases of the $a$-th wall, respectively. 
Using the variable $\psi$, the BPS equation 
for vector multiplet $(\ref{5DW-3.3})$ becomes 
\be
 \frac{1}{\z g^2}\p_y^2 {\rm Re}(\psi)
&\!\!\!=&\!\!\!
1-\sum_{A=1}^{N}\exp \Bigl( - 2{\rm Re}(\psi)+2m_A(y-Y_0)
+2\sum_{a=1}^{N-2}\a_A{}^a R_a \Bigl) \nonumber \\ 
&\!\!\!=&\!\!\!
1-\exp\left(-2{\rm Re} (\psi) +2W\right).\label{tong-17} 
\ee
Here, the explicit dependence on $y$ in Eq.(\ref{tong-17}) 
can be assembled by defining a function 
$W(y)$ as 
\be
W=\log\sum_{A=1}^{N}\exp \Bigl( 2m_A(y-Y_0)
+2\sum_{a=1}^{N-2}\a_A{}^a R_a \Bigl)\label{tong-50}.
\ee
To ensure that $\p_y \S =0$ at $y\ra\pm\infty$, 
we should impose boundary conditions for 
the above equation as 
\be
{\rm Re}(\psi) &\!\!\!\lra&\!\!\! 
m_1 (y-Y_0)+\sum_{a=1}^{N-2}\a_1{}^a R_a , \q y \ra +\infty, 
\\
{\rm Re}(\psi) &\!\!\!\lra&\!\!\! 
m_N (y-Y_0)+\sum_{a=1}^{N-2}\a_N{}^a R_a , \q y \ra -\infty. 
\label{tong-18}
\ee
To define an appropriate variable for the position of each wall, 
let us consider a configuration where 
only two adjacent exponential terms in the sum of Eq.(\ref{tong-17}) 
are large and the others are negligible, then 
the function $W$ has a profile for a single wall. 
Therefore it is natural to define the 
position $y_A$ of the $A$-th wall as  
\begin{eqnarray}
&\!\!\!&\!\!\!
\exp\left(2m_A (y_A-Y_0)
+2\sum_{a=1}^{N-2}\a_A{}^a R_a\right)=  
\exp\left(2m_{A+1} (y_A-Y_0)+2\sum_{a=1}^{N-2}\a_{A+1}{}^a R_a\right)
\nonumber\\
&\!\!\!&\!\!\!
\quad \rightarrow \quad y_A=Y_0-\frac{\sum_{a=1}^{N-2}
\left(\a_{A+1}{}^a-\a_A{}^a\right) R_a}{m_{A+1}-m_A}.\label{WallPos}
\end{eqnarray}
This moduli parametrization $y_A$ has an intuitive 
meaning of the position of the $A$-th wall and 
$y_{A}-y_{A+1}$ corresponds to a distance between two walls, 
at least when the distance is large. 
Relations between $y_A$ and the relative positions $R_a$ 
defined in Eq.(\ref{tong-16}) are 
obtained by 
\begin{eqnarray}
 -\sum_{a=1}^{N-2}(\a_A{}^a-\alpha_1{}^a) R_a=
\sum_{B=1}^{A-1}\left(m_{B+1}-m_{B}\right)(y_{B}-Y_0),
\quad A=2, \cdots, N.
\end{eqnarray}     
By choosing $A=N$, we find 
\begin{eqnarray}
 Y_0=\frac1{m_N-m_1}\left[\sum_{B=1}^{N-1}\left(m_{B+1}-m_{B}\right)y_{B}
+\sum_{a=1}^{N-2}(\alpha_N^a-\alpha_1^a)R_a\right], 
\end{eqnarray}
which becomes the center of mass coordinate when $\alpha_N^a=\alpha_1^a$.  
Because of the translational invariance, ${\rm Re}(\psi)$ 
should be a function of real variables $y-Y_0$ and $R_a$. 
From now on, we will take $Y_0=0$ unless otherwise stated.


Now we will present a series of exact solutions for single and 
double walls for finite values of gauge coupling. 
To compare single and multi-wall solutions, 
let us assign the following mass parameters 
for hypermultiplets for double walls ($N=3$) : 
\be
m^A = (m,0,-m) .
\label{eq:mass-double-wall}
\ee
This convention is intended to make the total energy density 
of the double wall to be identical to the single wall, so 
that the double wall situation can be most naturally 
compared to the single wall situation with the 
same energy density (tension) 
in the coincident limit of two walls, 
since the mass parameters of the single wall is assigned to be 
$m^A = (m,-m)$ in Eq.(\ref{eq:single-wall-mass}), and 
the total energy density is just given by the 
difference of the two extreme masses as given in 
Eq.(\ref{eq:total-tension}). 
For the moduli parameter, we use 
$y_1-y_2 = R^1 \equiv R$ and 
choose the $\alpha_A{}^1$ as 
\be
\alpha_A{}^1=(0, m/2, 0), 
\qquad 
2 \alpha_A{}^aR_a=2 \alpha_A{}^1R_1=(0, mR, 0), 
\label{eq:alpha}
\ee
since the rank of the matrix $\alpha_A{}^a$ is $N-2$. 
This relative distance appears only in multiple wall, 
but not in the single wall. 

The function $W\equiv W_{\rm single}$ for the single wall case becomes 
\be
W_{\rm single}\equiv \frac{1}{2}\log (\e^{2my}+\e^{-2my}). 
\ee
Similarly the function for the double wall case 
$W\equiv W_{\rm double}$ is found to be 
\be
W_{\rm double}\equiv \frac{1}{2}\log (\e^{2my}+\e^{-2my}+\e^{m R }).
\ee
The solvable cases of finite gauge coupling are found to be 
\be
g^2\z \equiv \frac{8m^2}{k^2},\quad 
k=0,2,3,4,
\label{eq:solvable-coupling}
\ee
where the mass parameter $m$ is defined 
for single wall case in Eq.(\ref{eq:single-wall-mass}), and double 
wall case in Eq.(\ref{eq:mass-double-wall}). 
We will denote the single wall solution as $S_{k}(m)$ and 
the double wall solution as $D_{k}(m)$ 
with the coupling defined by (\ref{eq:solvable-coupling}) 
with $k$ and the mass parameter $m$.

Let us list the exact solutions that we are able to obtain 
\be
&\!\!\!&\!\!\!S_0 (m)\q:\q 
\Re\psi =
\frac{1}{2}\log (\e^{2my}+\e^{-2my})\label{5DW-3.25},\\
&\!\!\!&\!\!\!
S_2 (m)\q:\q 
\Re\psi =\log (\e^{my}+\e^{-my}) \label{5DW-3.26},\\
&\!\!\!&\!\!\!
S_3 (m)\q:\q 
\Re\psi =
\frac{3}{2}\log (\e^{\frac{2}{3}my}+\e^{-\frac{2}{3}my})
\label{5DW-3.27},\\
&\!\!\!&\!\!\!
S_4 (m)\q:\q 
\Re\psi =\log (\e^{my}+\e^{-my}+\sqrt{6})\label{5DW-3.28},\\
&\!\!\!&\!\!\!
D_{0} (m)\q:\q 
\Re\psi =\frac{1}{2}\log (\e^{2my}+\e^{-2my}+\e^{mR})
\label{5DW-3.18},\\
&\!\!\!&\!\!\!D_{4} (m)\q:\q 
\Re\psi =\log (\e^{my}+\e^{-my}+\sqrt{6+\e^{mR}}) . 
\label{5DW-3.19}
\ee
\begin{figure}[thb]
\begin{center}
\leavevmode
\scalebox{1.0}{\includegraphics{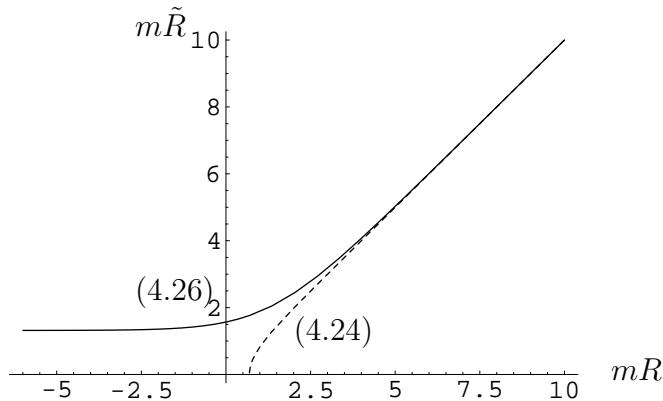}}
\begin{picture}(1,100)(0,0)
\put(-180,140){$m\tR$}
\put(-120,25){$(\ref{5DW-3.20})$}
\put(-180,40){$(\ref{5DW-3.21})$}
\put(0,10){$mR$}
\end{picture}
\vspace{0.5cm}
\caption{ Relation between $R$ 
defined in Eq.(\ref{tong-16}) and $\tR$. 
The solid line represents the relation 
defined in Eq.(\ref{5DW-3.21}) 
for the finite coupling solution $D_4(m)$ in 
Eq.(\ref{5DW-3.19}). 
The dashed line represents the relation 
defined in Eq.(\ref{5DW-3.20}) 
for the infinite coupling solution $D_0(m)$ in 
Eq.(\ref{5DW-3.18}). 
 }
\label{fig:R-tR-relation}
\end{center}
\end{figure}
It is interesting to observe that the double wall solution 
$D_{0} (m)$ in Eq.(\ref{5DW-3.18}) 
in the case of infinite coupling 
can be rewritten as a superposition of two  
single walls placed apart by a distance $\tilde R$ 
\be
\Re\psi(y)  
=\frac{1}{2}\log (\e^{m(y-\frac{\tR}{2})}
+\e^{-m(y-\frac{\tR}{2})})
+\frac{1}{2}\log (\e^{m(y+\frac{\tR}{2})}
+\e^{-m(y+\frac{\tR}{2})}).
\ee
This 
parameter $\tilde R$ can be regarded as another 
choice of a moduli parameter and is related to $R$ 
defined in Eq.(\ref{tong-16}) 
\be
\e^{m\tR}+\e^{-m\tR}=\e^{mR}\q\Ra\q 
m\tR = \log \Bigl( \frac{\e^{mR}
+\sqrt{\e^{2mR}-4}}{2} \Bigl), 
\label{5DW-3.20}
\ee
which is illustrated in Fig.\ref{fig:R-tR-relation}.
Unfortunately the new choice of the 
moduli parameter $\tilde R$ becomes pure 
imaginary and loses an intuitive meaning as 
the distance between the two walls 
when $R < \log 2/m$. 
This situation has some similarity to a moduli 
parameter for a model of BPS double wall 
in an ${\cal N}=1$ SUSY Wess-Zumino model in four-dimensions 
\cite{SV}. 
We can also rewrite the double wall solution 
$D_{4} (m)$ in Eq.(\ref{5DW-3.19}) 
in the case of the finite coupling 
as a superposition of two 
single walls placed apart by a distance $\tilde R$ 
\be
\Re\psi(y) 
=\log (\e^{\frac{m}{2}(y-\frac{\tR}{2})}
+\e^{-\frac{m}{2}(y-\frac{\tR}{2})})
+\log (\e^{\frac{m}{2}(y+\frac{\tR}{2})}
+\e^{-\frac{m}{2}(y+\frac{\tR}{2})}).
\ee
The new moduli parameter $\tilde R$ is related to $R$ 
in this case as 
\be
\e^{\frac{m\tR}{2}}+\e^{-\frac{m\tR}{2}}
=\sqrt{6+\e^{mR}}\q\Ra\q 
m\tR =
2\log \Bigl( \frac{\sqrt{6+\e^{mR}}
+\sqrt{2+\e^{mR}}}{2} \Bigl). 
\label{5DW-3.21}
\ee
In this case of the finite coupling 
$D_4(m)$ in Eq.(\ref{5DW-3.19}), 
the parameter $\tilde R$ takes only positive real values, 
whereas $R$ takes positive as well as negative values as 
illustrated in  Fig.\ref{fig:R-tR-relation}.
In both cases  of infinite and finite coupling, 
both $R$ and $\tilde R$ have an intuitive meaning 
of relative distance between the wall, as long as 
the distance is large : 
$\tR \rightarrow R$, for $R \rightarrow \infty$. 
Therefore $\tilde R$ in this case gives an 
intuitively nicer 
parametrization for the relative distance between 
the two walls.

\vspace{1cm}
\begin{figure}[htb]
\begin{center}
\leavevmode
\epsfxsize=6cm
\epsfysize=4cm
\begin{picture}(10,100)(150,10)
 \hspace{-15ex}
\epsfbox{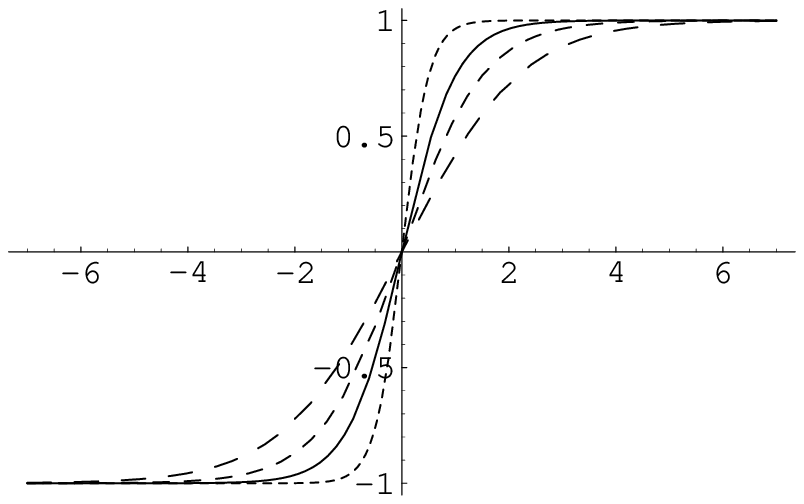}
\put(-10,60){$my$}
\put(-160,90){$S_{0}$ }\multiput(-140,93)(3,0){7}{\line(1,0){1.5}}
\put(-160,80){$S_{2}$ }\multiput(-140,83)(1,0){1}{\line(1,0){20}}
\put(-160,70){$S_{3}$ }\multiput(-140,73)(6,0){4}{\line(1,0){3}}
\put(-160,60){$S_{4}$ }\multiput(-140,63)(9,0){3}{\line(1,0){4.5}}
\put(-90,120){$\S$}
\put(-90,-15){a) }
\end{picture}
\epsfxsize=6cm
\epsfysize=4cm
\label{fig4-1-1}
\begin{picture}(20,100)(-30,10)
 \hspace{-5ex}
\epsfbox{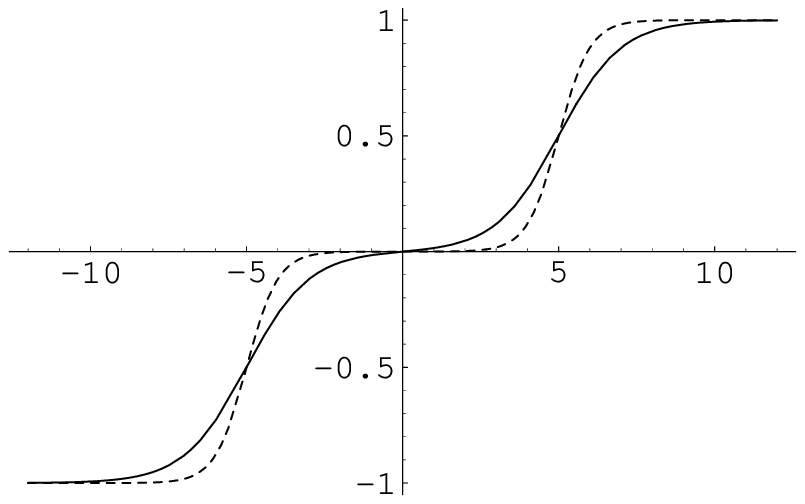}
\put(-10,60){$my$}
\put(-160,90){$D_{0}$ }\multiput(-140,93)(3,0){7}{\line(1,0){1.5}}
\put(-160,80){$D_{4}$ }\multiput(-140,83)(0,0){1}{\line(1,0){20}}
\put(-90,120){$\S$}
\put(-90,-15){b) }
\end{picture}
\end{center}
\vspace{0.5cm}
\caption{ Comparison of the vector multiplet scalar 
$\Sigma$ as a function of $my$ 
for exact solutions with various gauge couplings. 
a) Single wall solutions $S_0(m)$ (dotted line), 
$S_2(m)$ (solid line), $S_3(m)$ (short dashed line), 
and $S_4(m)$ (dashed line). 
b) Double wall solutions $D_0(m)$ (dotted line), and 
$D_4(m)$ (solid line). 
}
\label{fig4-1-2}
\end{figure}
\vspace{1cm}

\begin{figure}[htb]
\begin{center}
\leavevmode
\epsfxsize=6cm
\epsfysize=4cm
\begin{picture}(10,100)(150,10)
 \hspace{-15ex}
\epsfbox{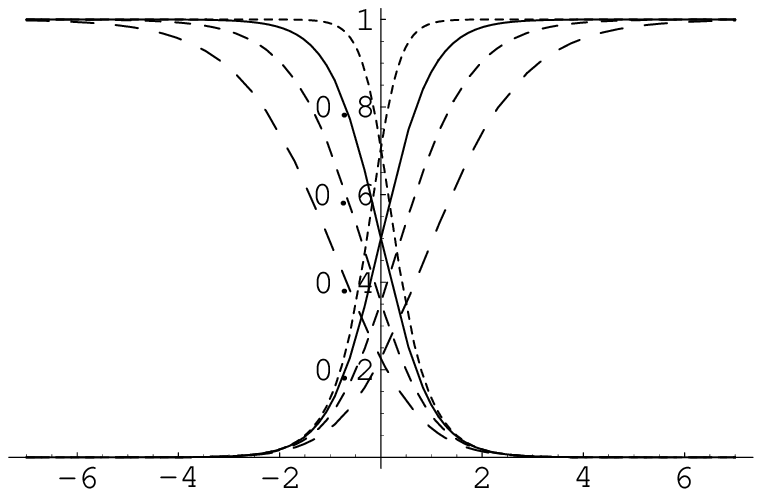}
\put(-160,90){$S_{0}$ }
\multiput(-140,83)(0,0){1}{\line(1,0){20}}
\put(-160,80){$S_{2}$ }
\multiput(-140,93)(3,0){7}{\line(1,0){1.5}}
\put(-160,70){$S_{3}$ }
\multiput(-140,73)(6,0){4}{\line(1,0){3}}
\put(-160,60){$S_{4}$ }
\multiput(-140,63)(9,0){3}{\line(1,0){4.5}}
\put(-10,15){$my$}
\put(-90,120){$H^{1A}$}
\put(-90,-15){a) }
\end{picture}
\epsfxsize=6cm
\epsfysize=4cm
\begin{picture}(20,100)(-30,10)
 \hspace{-5ex}
\epsfbox{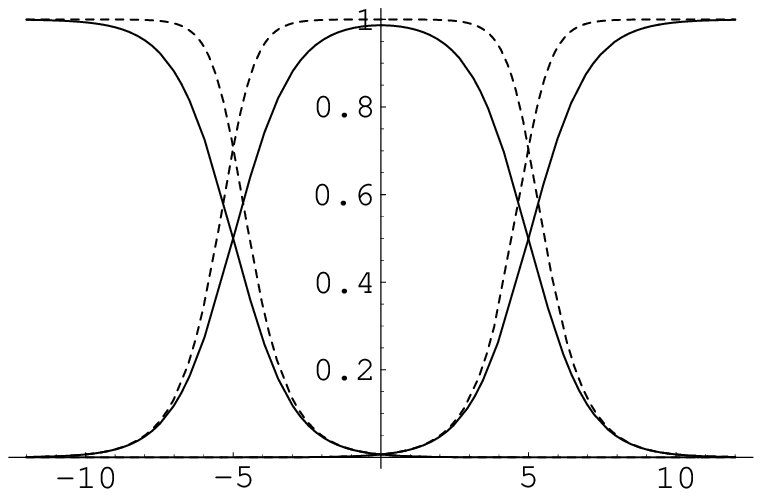}
\put(-180,90){$D_{0}$ }
\multiput(-160,93)(3,0){7}{\line(1,0){1.5}}
\put(-180,80){$D_{4}$ }
\multiput(-160,83)(0,0){1}{\line(1,0){20}}
\put(-10,15){$my$}
\put(-90,120){$H^{1A}$}
\put(-90,-15){b) }
\end{picture}
\vspace{0.5cm}
\caption{ Comparison of the hypermultiplet scalars 
$H^{1A}$ as a function of $my$ 
for exact solutions with various gauge couplings. 
a) Single wall solutions $S_0(m)$ (dotted line), 
$S_2(m)$ (solid line), $S_3(m)$ (short dashed line), 
and $S_4(m)$ (dashed line). 
b) Double wall solutions $D_0(m)$ (dotted line), and 
$D_4(m)$ (solid line). 
 }
\label{fig4-2-1}
\end{center}
\end{figure}

\begin{figure}[htb]
\begin{center}
\leavevmode
\epsfxsize=6cm
\epsfysize=4cm
\begin{picture}(10,100)(150,10)
 \hspace{-15ex}
\epsfbox{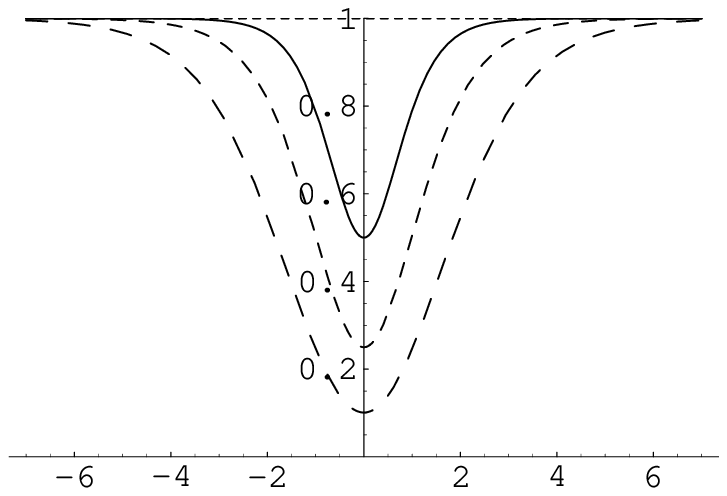}
\put(-160,90){$S_{0}$ }
\multiput(-140,83)(0,0){1}{\line(1,0){20}}
\put(-160,80){$S_{2}$ }
\multiput(-140,93)(3,0){7}{\line(1,0){1.5}}
\put(-160,70){$S_{3}$ }
\multiput(-140,73)(6,0){4}{\line(1,0){3}}
\put(-160,60){$S_{4}$ }
\multiput(-140,63)(9,0){3}{\line(1,0){4.5}}
\put(-10,15){$my$}
\put(-90,120){$\mathcal{V}$}
\put(-90,-15){a) }
\end{picture}
\epsfxsize=6cm
\epsfysize=4cm
\begin{picture}(20,100)(-30,10)
 \hspace{-5ex}
\epsfbox{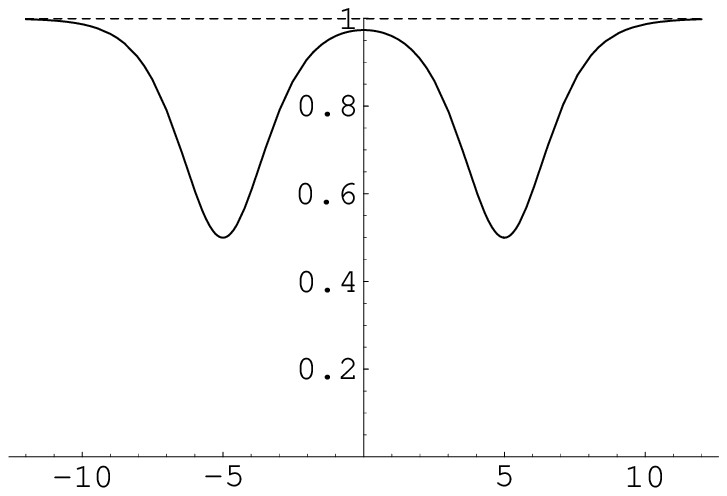}
\put(-180,90){$D_{0}$ }
\multiput(-160,93)(3,0){7}{\line(1,0){1.5}}
\put(-180,80){$D_{4}$ }
\multiput(-160,83)(0,0){1}{\line(1,0){20}}
\put(-10,15){$my$}
\put(-90,120){$\mathcal{V}$}
\put(-90,-15){b) }
\end{picture}
\vspace{0.5cm}
\caption{ Comparison of potential 
$\mathcal{V} \equiv {V \over 2g^2 \zeta}$ 
normalized by the asymptotic value 
($y\rightarrow \pm \infty$) 
as a function of $my$ 
for exact solutions with various gauge couplings. 
a) Single wall solutions $S_0(m)$ (dotted line), 
$S_2(m)$ (solid line), $S_3(m)$ (short dashed line), 
and $S_4(m)$ (dashed line). 
b) Double wall solutions $D_0(m)$ (dotted line), and 
$D_4(m)$ (solid line). 
}
\label{fig:potential}
\end{center}
\end{figure}

It is interesting to examine how these double wall 
solutions behave, in the limit of coincident walls 
and in the limit of asymptotically far apart walls. 
By smoothly changing the moduli parameter $R$, 
we find the following limiting 
behaviors which can be expressed 
symbolically as 
\be
S_0(m)\st{R\ra -\infty
}{\longleftarrow}
&D_{0}(m)&\st{R\ra +\infty
}
{\longrightarrow}
S_0 \left(\frac{m}{2}\right)\oplus 
S_0 \left(\frac{m}{2}\right),\\
S_4(m)\st{R\ra -\infty
}
{\longleftarrow}
&D_{4}(m)&\st{R\ra +\infty
}
{\longrightarrow}
S_2 \left(\frac{m}{2}\right)\oplus 
S_2 \left(\frac{m}{2}\right).
\ee  
These limiting behaviors apply not only for the 
$\psi(y)$, but also 
for all the physical quantities. 
Note that the sum of the numbers $k$ is preserved, 
corresponding to 
the fact that the total tension is a conserved topological 
invariant for a given boundary condition. 
We can find $\S,H^{1A}$ and the potential $V$ 
corresponding to these exact single 
and double wall solutions (\ref{5DW-3.25})--(\ref{5DW-3.19}) 
by using (\ref{tong-13}), (\ref{tong-15}) 
and (\ref{5DW-1.20}). 
We illustrate and compare various single and double wall 
solutions for scalar $\Sigma(y)$ of vector multiplet in 
Fig.\ref{fig4-1-2},  
for hypermultiplet scalars $H^{1A}(y)$ for $A=1,2$ 
(single wall) and 
$A=1, 2, 3$ (double wall) in 
Fig.\ref{fig4-2-1} and for potential $V(y)$ in 
Fig.\ref{fig:potential}.

\vspace{1cm}
\vspace{1cm}

\subsection{{ Moduli Dynamics of Two Domain Walls}}

The low-energy effective Lagrangian for massless field has been studied 
systematically, especially in the case of infinite gauge coupling 
\cite{Tong}. 
It has been proposed to use an expansion in inverse powers of gauge 
coupling, since the infinite gauge coupling case was the only 
explicitly known solution at that time. 
We would like to derive the low-energy effective Lagrangian 
for massless field in the case of finite gauge coupling, 
using our exact solutions. 
We will compare it to the previous result of infinite gauge coupling, 
and show an instability of an expansion in inverse powers 
of gauge coupling. 

Since the overall position and phase become just free massless 
fields \cite{Tong}, 
we shall concentrate on the relative distance $R_a$ and 
relative phase $\theta_a$ of multiple walls defined in 
Eq.(\ref{tong-16}). 
These moduli parameters represent positions in $y$ and 
in internal space of $U(1)$ of multiple walls. 
We have obtained mode functions of massless Nambu-Goldstone fields 
by differentiating the background in terms of these moduli 
parameters in Sect.\ref{sc:fluctuation}. 
Here we are interested in not only the mass spectrum 
but also the entire effective Lagrangian at low energies. 
To find out the low-energy effective Lagrangian, 
a more systematic method by Manton \cite{Manton} 
is useful. 
In this method, we need to promote the parameters 
of the solution (moduli) to be functions of the world volume 
coordinates $x^\mu$, namely four-dimensional fields. 
Assuming the variation in the world volume coordinates $x^\mu$ 
to be weak, we can obtain all the nonlinear dynamics containing 
smallest number of derivatives (two derivatives). 
This well-defined procedure allows us to obtain the nonlinear 
kinetic term of the massless fields. 
Since we are considering the massless Nambu-Goldstone fields, 
there should be no additional potential terms without 
derivatives. 
We can keep only $t\equiv x^0$ dependence of these fields, 
since 
the Lorentz invariance in the world volume allows to 
recover the entire 
dependence on $x^\mu$ \cite{Tong}. 
Effective Lagrangian $\Lag_{\rm eff}$ 
of moduli fields is found to be given by 
integrating over the Lagrangian of the fundamental theory 
evaluated by the background field with the $t$ 
dependent moduli $r_a(t)=R_a(t)+i\theta_a(t)$ \cite{Tong}
\be
\Lag_{\rm eff} 
=\int_{-\infty}^{\infty} dy \sum_{A=1}^N |H^{1A}|^2 
(\dpsi + m_A \dot{y}_0-\a_A^a \dr^a )
( m_A \dot{y}_0^\dag-\a_A^a \dr^\dag_a  ).
\ee

Let us take two wall case ($N=3$) for concreteness. 
We discard terms for the overall position and phase, and 
omit the subscript $a$, since there is only one set of 
relative moduli $R, \theta$ now.  
We will be interested in low-energy effective Lagrangian 
for the relative distance $R$ and 
the relative phase $\theta$ of two walls. 
To obtain the dependence on the relative phase 
$\theta$, we need to find the imaginary part of $\psi(y)$ 
besides the real part that we have determined. 
The BPS equation (\ref{tong-15}) 
shows that the imaginary part of $\psi(y)$ is given by 
integrating the extra dimension component 
of the gauge field : 
$\partial_y {\rm Im}\psi(y)=W_4(y)$. 
As shown in the case of the single wall 
in Sect.\ref{sc:fluctuation}, 
we have, near the walls, 
a nontrivial field strength $F_{\mu y}(W) \not =0$ 
proportional to the derivative of the moduli fields 
$\dot \theta$. 
Then the extra dimension component of the vector 
potential is determined by the field equation for the 
vector potential 
\cite{Tong}. 
The most convenient gauge 
 to make the remaining SUSY 
manifest is $W_\mu=0$ and $W_4 \not=0$. 

We find from the equation of motion 
that the imaginary part (combined with the real part) 
satisfies 
\begin{equation}
{\partial^2 \dot{\psi} \over \partial y^2}=
2g^2 \zeta \left(\dot{\psi} 
- \dot{r} {\partial W(r+r^\ast) \over \partial r}
\right) e^{-\psi-\psi^*-2W}. 
\end{equation}
Because of the boundary condition $\psi=0$ at $y=\pm \infty$, 
the solutions of the real and imaginary parts 
are proportional to $\dot R$ and $\dot \theta$ 
with the same coefficient which is a function of $R$ only 
\begin{equation}
{\rm Re} \dot \psi 
=\dot R {\partial \over \partial R} 
{\rm Re}\psi(y, R), 
\quad 
{\rm Im}\dot \psi 
=\dot \theta {\partial \over \partial R} 
{\rm Re}\psi(y, R), 
\end{equation}
implying 
$\dot W_4=
\partial_y {\partial \over \partial R} 
{\rm Re}\psi(y, R) \dot{\theta}$. 
The moduli 
field $r\equiv R+i\theta$ is 
a complex scalar of a chiral scalar field, 
taking values \cite{Tong} 
$R\in \mathbf{R}$ and $\th \in [0,4\pi/m)$. 
We see that the ${\cal N}=1$ SUSY in 
four dimensions is now 
manifest. 
Moreover, the K\"ahler metric ${\cal K}_{r r^*}$ 
of the low-energy effective Lagrangian $\Lag_{\rm eff}$ 
is given in terms of 
$F(R)$ which is 
a function of $R$ only 
\be
\Lag_{\rm eff} =
\left(\dR^2 + \dth^2 \right)
{\cal K}_{r r^*}(r, r^*), 
\ee
\be
{\cal K}_{r r^*}(r, r^*) =
\frac{1}{4}m\z F(mR)
=
\frac{1}{4}m\z F
\left(m{ r + r^* \over 2}\right). 
\ee
The function $F(R)$ is most conveniently evaluated 
by 
taking $\theta(t)=0$ 
and by 
writing hypermultiplets $H^{1A}$ in terms of 
the function $\psi(y)$ in Eq.(\ref{tong-13}) and 
using the choice $\alpha_A$ in Eq.(\ref{eq:alpha}) 
\be
\frac{1}{4}m\z \dR^2 F(mR) 
&\!\!\!=&\!\!\!
\int_{-\infty}^{\infty} dy \sum_{A=1}^3 |H^{1A}|^2 
(-\Re \dpsi\a_A \dR +\a_A^2  \dR^2 ) 
\nonumber \\
&\!\!\!=&\!\!\!
\frac{1}{4}m\z \dR^2 
{d \over dR}\int_{-\infty}^{\infty} dy 
{\rm e}^{-2\psi(y)+mR}.
\label{eq:kaehler-metric}
\ee

The K\"ahler metric ${\cal K}_{r r^*}$ gives 
equations of motion for the relative 
distance $R$ and for the relative phase $\theta$ 
\be
\ddot{R} &\!\!\!=&\!\!\!
-\frac{1}{2}m(\log F )^{'}(\dR^2 - \dth^2 ),
\label{nlsm1}\\
\ddot{\th}&\!\!\!=&\!\!\!-m(\log F)^{'}\dR\dth,
\label{nlsm2}
\ee
respectively, 
where $'$ means $\p /\p z, \; z\equiv mR$. 
These equations of motion are real 
and imaginary parts of a single equation due to the 
complex structure of the ${\cal N}=1$ SUSY : 
\be
\ddot{r}=-\frac{1}{2}m(\log F)^{'}\dr^2,\q r\equiv R+i\th 
\label{nlsm3}
\ee
We observe that the above equations of motion admit 
that $\dot \theta=0$ is always a consistent solution. 
Therefore we can discuss the relative motion of the 
double wall with fixed relative phase consistently. 
On the other hand, in order to have a fixed $R$, 
the above equations of motion is consistent only if 
$\theta$ is also constant. 
Therefore we shall consider the relative motion 
assuming a constant relative phase $\theta$ in the 
following. 

Inserting our exact wall 
solution $D_{4}(m)$  
 with the finite gauge coupling in 
Eq.(\ref{5DW-3.19}) into Eq.(\ref{eq:kaehler-metric}), 
we obtain 
the corresponding K\"ahler metric $F(z=mR)$ 
for the finite gauge coupling 
\be
F(z)_{4}&\!\!\!=&\!\!\!\e^z\int_{-\infty}^{\infty} dt \frac{\e^t+\e^{-t}
+\frac{6}{\sqrt{6+\e^z}}}{(\e^t+\e^{-t} + \sqrt{6+\e^z})^3}
\nn\\
&\!\!\!=&\!\!\!\frac{\e^z}{(2+\e^z)^2}
\Bigl\{ \e^z -4 + \frac{24}{\sqrt{(2+\e^z )(6+\e^z)}}
\log \Bigl[ \frac{\sqrt{6+\e^z}
+\sqrt{2+\e^z}}{2} \Bigl] \Bigl\}. 
\ee
\vspace{0.8cm}
\begin{figure}[ht]
\begin{center}
\leavevmode
\scalebox{1.0}{\includegraphics{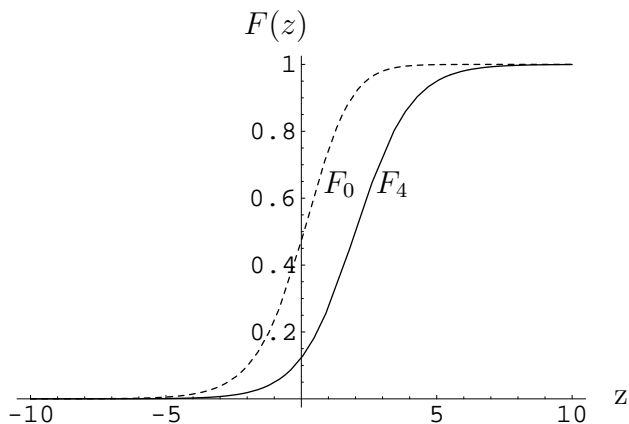}}
\begin{picture}(1,100)(0,0)
\put(-140,150){$F(z)$}
\put(-110,90){$F_{0}$}
\put(-90,90){$F_{4}$}
\put(0,10){z}
\end{picture}
\vspace{0.3cm}
\caption{ Comparison of K\"ahler metrics 
$F(z)_{0}$ for the infinite gauge coupling 
(dashed line) 
and $F(z)_{4}$ for the finite gauge coupling 
(solid line), 
as functions of $z=mR$.}
\label{fig5}
\end{center}
\end{figure}
For the $D_{0}(m)$ solution 
at the infinite coupling, 
the K\"ahler metric 
$F$ has been obtained 
 as 
\cite{Tong} 
\be
F(z)_{0}&\!\!\!=&\!\!\!
\frac{\e^z}{2}\int_{-\infty}^{\infty} dt 
\frac{\e^t+\e^{-t}}{(\e^t+\e^{-t}+\e^z )^2}
\nn\\
&\!\!\!=&\!\!\!
\frac{\e^z}{\e^{2z}-4}
\Bigl\{ \e^z + \frac{4}{\sqrt{\e^{2z}-4}}
\log \Bigl[ \frac{2}{\e^z + \sqrt{\e^{2z}-4}} \Bigl] 
\Bigl\}. 
\label{eq:kaehler0}
\ee
The metric is real and positive for the entire values 
of $-\infty < z < \infty$, in spite of the apparent 
singularity at ${\rm e}^{2z}=4$ \cite{Tong}. 
These K\"ahler metrics are illustrated and compared 
in Fig.\ref{fig5}. 
We see that there is a significant difference 
between them 
quantitatively, although general features are similar. 

\begin{figure}[htb]
\begin{center}
\leavevmode
\scalebox{1.0}{\includegraphics{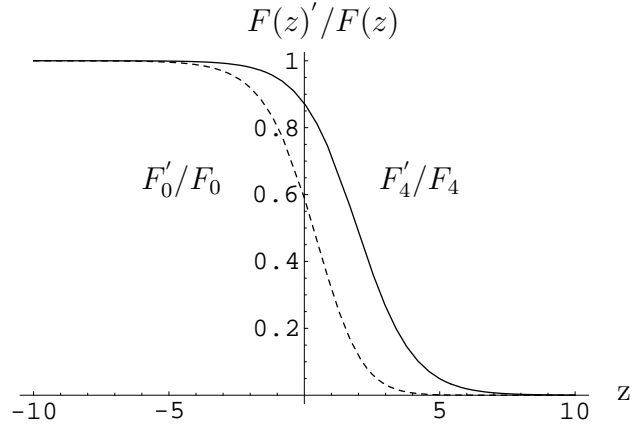}}
\begin{picture}(1,100)(0,0)
\put(-140,150){$F(z)^{'}/F(z)$}
\put(-180,90){$F_{0}^{'}/F_{0}$}
\put(-90,90){$F_{4}^{'}/F_{4}$}
\put(0,10){z}
\end{picture}
\vspace{0.1cm}
\end{center}
\caption{ Comparison of the ratio $F_{0}^{'}/F_{0}$ 
(dashed line) and $F_{4}^{'}/F_{4}$ (solid line).
}
\label{fig:ratio-forces}
\end{figure}

The physical significance of the quantitative difference 
can perhaps most easily appreciated by comparing 
the strengths of forces acting between the walls 
in the case of the finite coupling to that of 
the infinite coupling. 
In Fig.\ref{fig:ratio-forces}, 
we illustrate and compare the coefficients 
$F'/F$ of force in 
Eqs.(\ref{nlsm1})--(\ref{nlsm3}) for the 
infinite and finite coupling cases. 
We observe that the strength of the force for the 
finite coupling case 
is much larger than the infinite coupling case 
for large relative distance $R$.

\vspace{0.8cm}
\vspace{1cm}

To understand the quantitative difference of the 
strength of the forces at large values of $R$, 
let us evaluate the K\"ahler metric $F$ 
at asymptotic values of $R$. 
In the region of $R\ra -\infty$, 
both target space metrics $F(z)$ are 
flat and 
are related each other by a scale transformation. 
On the other hand, in the case of $R\ra +\infty$, 
our Lagrangian for the finite 
coupling becomes 
\be
\Lag_{4}
&\!\!\!=&\!\!\!
\frac{1}{4}m\z \Bigl[ 1-8\e^{-mR}+(12mR+28)\e^{-2mR}\Bigl] 
(\dR^2 + \dth^2 ). 
\ee
Consequently the equation of motion of the 
 relative distance $R$ for $D_4(m)$ solution at the 
finite gauge coupling 
is found to be 
\be
\ddot{R}&\!\!\!=&\!\!\! 
\Bigl[ -\e^{-mR}4m + \e^{-2mR}(12m^2 R-10m)\Bigl] 
(\dR^2 - \dth^2 ).
\label{eq:EOM-finite-coupling}
\ee
On the other hand, the 
asymptotic Lagrangian for infinite coupling 
at $R\rightarrow \infty$ has been found to be 
\cite{Tong} 
\be
\Lag_{0}
&\!\!\!=&\!\!\!
\frac{1}{4}m\z \Bigl[ 1-4(mR-1)\e^{-2mR}\Bigl] 
(\dR^2 + \dth^2 ), 
\ee
and the corresponding $R\rightarrow \infty$ limit of the 
equation of motion 
of the relative 
distance $R$ for $D_0(m)$ 
solution at the infinite coupling 
reads 
\cite{Tong} 
\be
\ddot{R}&\!\!\!=&\!\!\! 
-\e^{-2mR}(4m^2R-6m)(\dR^2 - \dth^2 ). 
\label{eq:EOM-infinite-coupling}
\ee
We see that for $\dR^2 > \dth^2$ 
there is always an attractive force operating between 
walls which are sufficiently far apart, 
irrespective of the strength of the gauge coupling. 
However, we find that the strength of the force 
for the finite 
coupling behaves as ${\rm e}^{-mR}$, which is much 
larger than that for the infinite coupling 
case of ${\rm e}^{-2mR}$. 
This explains the reason why we obtained much stronger 
force between the walls in the case of 
the finite coupling.

\subsection{{ Approximation in $1/g^2$ and 
Asymptotic Behaviors}}

Here we wish to discuss the previously proposed expansion 
in inverse powers of  gauge coupling to obtain finite 
coupling results \cite{Tong}. 
We also analyze the asymptotic behavior of the BPS wall 
solutions for generic values of gauge coupling 
to understand the reason why the power series approximation 
exhibits pathological behavior. 

We expand ${\rm Re}\psi$ in inverse powers of the gauge coupling : 
$a\equiv(2g^2\zeta)^{-1}=k^2/(4m)^2$ as 
\begin{equation}
{\rm Re}\psi=\sum_{n=0}^\infty a^n\psi_n.  
\end{equation}
By substituting the expansion to the equation 
(\ref{tong-17}), we obtain an expansion in 
power of $a$. 
The leading power $1/a$ comes only from the right-hand side 
and gives the result at the infinite coupling in 
Eq.(\ref{5DW-3.18}) : $\psi_0=W$. 
Using this result, all the successive powers 
can be solved iteratively
\begin{equation}
\sum_{n=0}^\infty a^n \psi_{n+1}=
\sum_{n=0}^\infty a^n {d^2 \psi_n \over dy^2}
-2\sum_{n=1}^{\infty}{(-2)^n a^n \over (n+1)!}
\left(\sum_{l=0}^{\infty} a^l \psi_{l+1}\right)^{n+1}.  
\end{equation}
Several lower order results are given explicitly by 
\begin{eqnarray}
 \psi_0&\!\!\!=&\!\!\!W,\quad \psi_1=W^{(2)},
 \quad\psi_2=W^{(4)}+(W^{(2)})^2,\nonumber\\
\psi_3&\!\!\!=&\!\!\!
W^{(6)}+4W^{(4)}W^{(2)}+2(W^{(3)})^2
+{4\over 3}(W^{(2)})^3, \nonumber\\
\psi_4&\!\!\!=&\!\!\!
W^{(8)}+6W^{(6)}W^{(2)}+12W^{(5)}W^{(3)}
+9(W^{(4)})^2+12W^{(4)} (W^{(2)})^2  \nonumber\\
&\!\!\!&\!\!\!+ 12(W^{(3)})^2 W^{(2)}
+4(W^{(2)})^4, 
\cdots,
\label{eq:power-series-approx}
\end{eqnarray}
where $W^{(n)}$ is defined by $ W^{(n)}\equiv d^n W/d y^n$. 

\begin{figure}[htb]
\begin{center}
\leavevmode
\scalebox{1.0}{\includegraphics{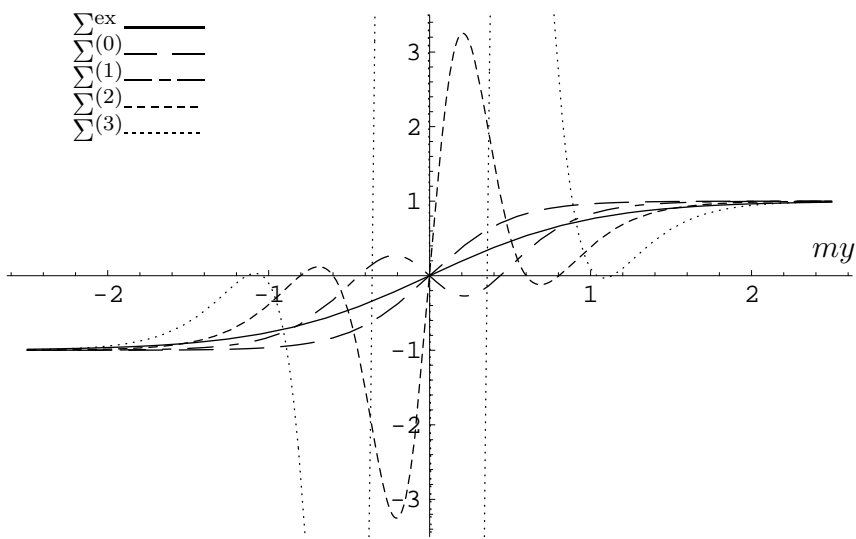}}
\begin{picture}(1,100)(0,0)
\put(-300,190){$\S^{\rm ex}$ }
\multiput(-280,193)(0,0){1}{\line(1,0){30}}
\put(-300,180){$\S^{(0)}$ }
\multiput(-280,183)(0,0){1}{\line(1,0){12}}
\multiput(-262,183)(0,0){1}{\line(1,0){12}}
\put(-300,170){$\S^{(1)}$ }
\multiput(-280,173)(0,0){1}{\line(1,0){10}}
\multiput(-267,173)(0,0){1}{\line(1,0){4}}
\multiput(-260,173)(0,0){1}{\line(1,0){10}}
\put(-300,160){$\S^{(2)}$ }
\multiput(-280,163)(5,0){6}{\line(1,0){2.5}}
\put(-300,150){$\S^{(3)}$ }
\multiput(-280,153)(3,0){10}{\line(1,0){1.0}}
\put(-20,106){$my$}
\end{picture}
\vspace{0.1cm}
\end{center}
\caption{ 
Comparison of approximations in inverse powers of 
gauge coupling to the exact solution $\Sigma^{\rm ex}(y)$ 
for the vector multiplet scalar 
$\Sigma (y)=\partial_y{\rm Re}\psi(y)$ with a finite 
gauge coupling 
$g^2\zeta=2m^2$. 
Approximation up to the $l$-th order is denoted as 
$\Sigma^{(l)}$. 
The zero-th order approximation (infinite gauge coupling) 
$\Sigma^{(0)}$ is represented by a dashed line. 
$\Sigma^{(1)}$ by a dash-dotted line, 
$\Sigma^{(2)}$ by a short dashed line, 
and 
$\Sigma^{(3)}$ by a dotted line. 
The mass parameter is set to $m=1$. 
}
\label{fig:sigma-power}
\end{figure}

First let us compare the power series approximation 
with our exact solution of the single wall $S_2(m)$ 
for the finite gauge coupling 
$k=2$. 
We illustrate the result of the power series 
expansion up to the $l$-th order 
$\Sigma^{(l)}(y)=\partial_y\psi^{(l)}(y)\equiv 
\sum_{n=0}^l a^n \partial_y\psi_n(y)$ 
by setting $a\equiv(2g^2\zeta)^{-1}=(2m)^{-2}$, 
and compare it with the exact solution 
$\partial_y\psi(y)$ 
in Fig.\ref{fig:sigma-power}. 
Near the wall, the approximation oscillates wildly 
and shows no indication of convergence. 
\vspace{0.8cm}

Similarly we can obtain the power series approximations 
for the K\"ahler metric $F$ 
for the relative position and phase for two walls. 
For concreteness, we will take the case of our exact 
solution $D_4(m)$ 
for the finite coupling $k=4$, and 
compare 
the exact K\"ahler metric 
$F(z)_4$ with a profile obtained by
the approximation in inverse powers of the gauge coupling. 
We can obtain the approximations 
up to the $l$-th order 
$F(z)_4^{(l)}$, 
by inserting the above 
expansion (\ref{eq:power-series-approx}) 
to the action (\ref{eq:kaehler-metric}) and by 
setting $a=(2g^2\zeta)^{-1}=m^{-2}$.  
\begin{figure}[hbt]
\begin{center}
\leavevmode
\epsfxsize=7cm
\epsfysize=5cm
\begin{picture}(200,200)(20,0)
\epsfbox{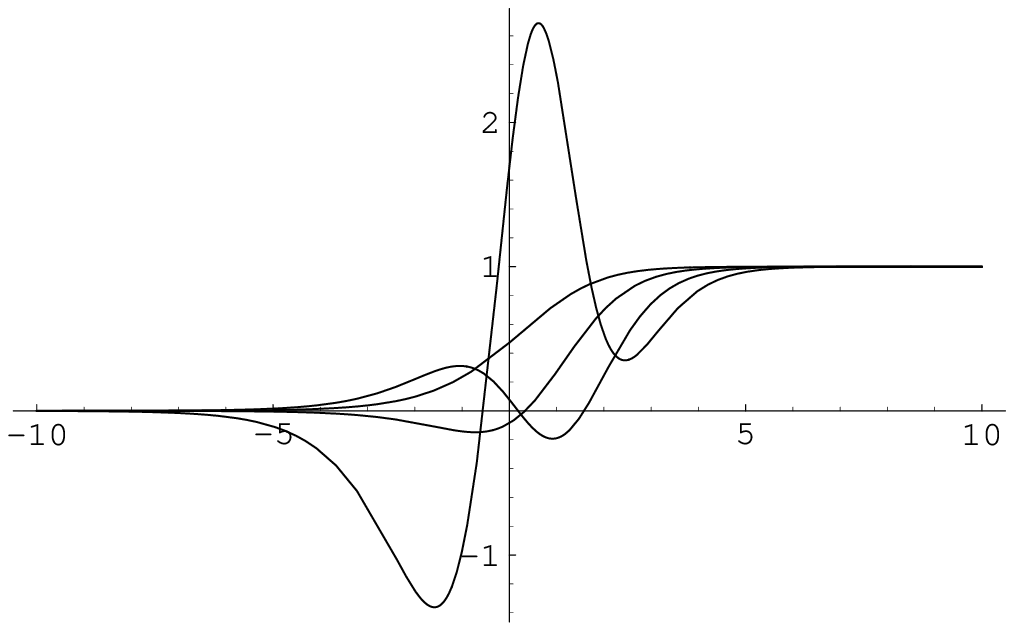}
\put(-110,145){$F(z)$}
\put(0,45){$z$}
\put(-100,75){\footnotesize $F_0$}
\put(-130,35){\footnotesize $F_4^{(1)}$}
\put(-85,35){\footnotesize $F_4^{(2)}$}
\put(-85,100){\footnotesize $F_4^{(3)}$}
\put(-85,-10){a)}
\end{picture}
\epsfxsize=7cm
\epsfysize=5cm
\begin{picture}(200,200)(0,-10)
\epsfbox{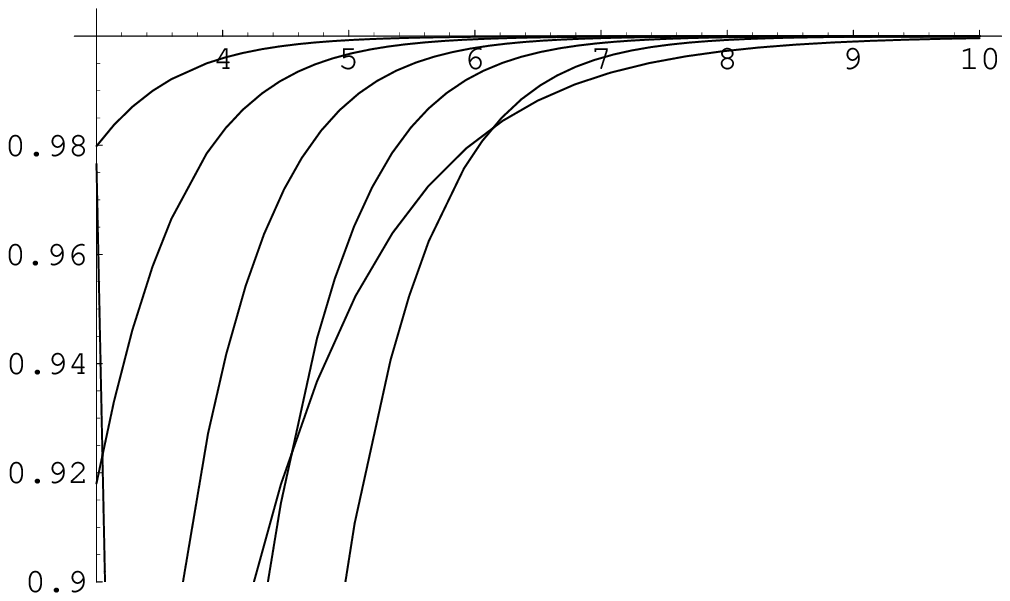}
\put(-200,145){$F(z)$}
\put(0,130){$z$}
\put(-90,110){\footnotesize $F_{4}$}
\put(-180,120){\footnotesize $F_0$}
\put(-180,90){\footnotesize $F_4^{(1)}$}
\put(-170,70){\footnotesize $F_4^{(2)}$}
\put(-150,70){\footnotesize $F_4^{(3)}$}
\put(-120,50){\footnotesize $F_4^{(4)}$}
\put(-120,-10){b)}
\end{picture}
\end{center}
\caption{Comparison of approximations in inverse powers of 
gauge coupling up to the $l$-th order 
$F_4^{(l)}(z)$ to the exact solution $F_4(z)$ 
for the K\"ahler metric 
of moduli fields as a function of 
$z\equiv mR$. 
a) Near the region of coincident walls. 
b) Asymptotic region $y = mR \rightarrow \infty$. 
}
\label{figure.8}
\end{figure}
In Fig.\ref{figure.8}a), 
we compare the power series 
approximation up to various orders 
starting from the zero-th order $F_0$ (infinite gauge 
coupling case) up to the third order approximations 
$F_4^{(3)}$. 
We see wild oscillations near the wall. 
As the order of approximation increases, this oscillatory 
behavior becomes wilder, without any indication of dying out. 
If we abandon to describe the region near the wall, 
we can examine the approximation in the asymptotic 
regions away from the wall. 
We also illustrate 
the K\"ahler metric $F(z)$ for larger values of 
$z=m R\rightarrow \infty $ in 
Fig.\ref{figure.8}b). 
Even there, there is no indication for the series 
to converge to our exact result. 
Actually, we can make the discrepancy more quantitative. 
For instance let us take the power series 
approximation up to the fourth order $F(mR)_k^{(4)}$ 
for the function $F(m R)$ 
with an arbitrary coupling $a=(2g^2\zeta)^{-1}=(k/(4m))^2 $. 
Its asymptotic behavior 
at $R \rightarrow \infty$ is given by  
\begin{eqnarray}
 F(mR)_k^{(4)}&\!\!\!=&\!\!\!1+\left(-4(1+4m^2a+16m^4a^2
 +64m^6a^3+256m^8a^4)mR
\right.\nonumber \\ 
&\!\!\!&\!\!\! \left. {}\qquad +4+24m^2a +128m^4a^2
+\frac{2176}3m^6a^3+\frac{58112}{15}m^8a^4
+O(a^5)\right)e^{-2mR}\nonumber \\
&\!\!\!&\!\!\!+O(e^{-4mR}),
\end{eqnarray}
whereas  the exact form of the function with the finite coupling at $R\rightarrow \infty$ is 
\begin{eqnarray}
 F(mR)_4&\!\!\!=&\!\!\!1-8e^{-mR}+(12mR+28)e^{-2mR}+O(e^{-4mR}). 
\end{eqnarray}
It is now clear that the power series approximation 
$F(mR)_4^{(l)}, l \rightarrow \infty$ can never converge 
to the function $F(mR)_4$.  
Therefore we conclude that the approximation by an expansion 
in inverse powers of the gauge 
coupling is pathological, and may not be suitable 
to extract physical quantities correctly. 

Finally we analyze 
the 
asymptotic behavior of the BPS solution for 
arbitrary finite gauge coupling. 
This can be worked out even without exact solutions 
and may provide an alternative approximation scheme, 
rather than the expansion in inverse powers of gauge 
coupling. 
Let us take single wall case with two hypermultiplets 
for simplicity. 
Since BPS solutions are essentially real when the moduli 
parameters are constants (not fields), 
we can ignore Im$\psi(y)$ and obtain 
the BPS equation as 
\begin{equation}
 {d^2\psi \over dy^2}
 =g^2\zeta \left(1-e^{-2\psi }(e^{2 my}+e^{-2my})\right)
\end{equation}
Let us parametrize the gauge coupling in terms of $k$ 
which is no longer an integer here : 
$g^2\zeta =8m^2/k^2$. 
We wish to devise an approximation scheme valid for 
large values of $z\equiv my$. 
Since the vector multiplet scalar 
$\Sigma = \partial_y \psi(y)$ 
should approach to a constant asymptotically, 
the BPS equation dictates that $\psi(y)$ should 
approach $\pm my$ at $y \rightarrow \pm \infty$. 
We expand the remaining subleading order terms in a series 
which will be determined successively 
as solutions of iterative 
equations 
\begin{eqnarray}
\psi(y) =m y +\frac12\sum_{n=1}^\infty \varepsilon _n(y),
\end{eqnarray}
where we can assume 
that the remaining terms should vanish 
asymptotically, since we can 
absorb a possible constant term into a shift of 
$y$ 
\begin{eqnarray}
 \varepsilon _n(y)\rightarrow 0
 \quad  {\rm as~}y\rightarrow 0, 
 \qquad \varepsilon _1(y)\gg \varepsilon _2(y)\gg \cdots. 
\end{eqnarray}
By substituting the expansion to the BPS equation, 
we obtain 
\begin{eqnarray}
 \frac12\sum_{n=1}^\infty {d^2\varepsilon _n\over dy^2}&=&
g^2\zeta \left(1-e^{-\sum_{n=1}^\infty \varepsilon _n}
(1+e^{-4my})\right).
\end{eqnarray}
We can rewrite this equation with $z\equiv  my$ and 
$g^2\zeta =8m^2/k^2 $  
\begin{eqnarray}
\sum_{n=1}^\infty {d^2 \varepsilon _n\over d z^2}
&=&{16\over k^2}
\left(1-e^{-\sum_{n=1}^\infty 
\varepsilon _n}(1+e^{-4z})\right),
\nonumber \\
&=&{16\over k^2}
\left(\varepsilon _1+\varepsilon _2
-\frac12\varepsilon ^2_1+ \cdots
-e^{-4z}(1-\varepsilon _1+\cdots)\right). 
\end{eqnarray}
The first order approximation is determined by the following 
equation 
\begin{equation}
{d^2\varepsilon_1 \over dz^2}
={16\over k^2}\left(\varepsilon_1 -e^{-4z}\right),
\end{equation}
which yields a general solution with an arbitrary 
integration constant $c_1$ 
\begin{eqnarray}
\varepsilon_1 =2c_1e^{-\frac{4z}{k}}
+{1\over 1-k^2}e^{-4z}, 
\end{eqnarray}
where we have used the boundary condition $\varepsilon_1 
\rightarrow 0$ at $y \rightarrow \infty$ 
to eliminate another possible integration constant. 
Therefore the asymptotic behavior of the 
first order approximation $\varepsilon _1(y)$ depends 
on the strength of the gauge coupling constant 
$g^2={8m^2 \over k^2 \zeta}$ as 
\begin{eqnarray}
\varepsilon _1=
 \left\{\begin{array}{ccc}
  2c_1e^{-\frac{4z}{k}}&:& k>1, c_1>0  \\
  2(z-z_0)e^{-4z}&:& k=1\\
  {1\over 1-k^2}e^{-4z}&:&0<k<1
 \end{array}\right. .
\end{eqnarray}
The wall solution requires that the vacuum value $m$ 
is reached from below : 
$\Sigma (y) =\partial_y \psi(y) < m$ 
as $y \rightarrow \infty$, 
This implies $\varepsilon _1(y)>0$. 
Therefore we obtain up to the first order approximation 
\begin{eqnarray}
 \psi \rightarrow 
 \left\{\begin{array}{ccc}
  my+c_1e^{-\frac4k my}&:& k>1 \\
  my+m(y-y_0)e^{-4 my}&:& k=1\\
  my+{1\over 1-k^2}e^{-4my}&:&0<k<1
 \end{array}\right.
\end{eqnarray}
This result can be regarded as the origin of 
a different asymptotic behavior for the finite coupling 
exact solution ($k=2$ case) compared to 
the infinite coupling solution ($k=0$ case). 
It is interesting to observe that the critical case 
of $k=1$ has an exceptional asymptotic behavior 
corresponding to degenerate exponents. 
This complication 
may be related to the fact that we have not yet 
succeeded to obtain an exact solution in that case. 

It is now straightforward to extend our approximation scheme 
to more general multi-wall cases. 
Since our approximation scheme from asymptotic region 
is applicable to any values of gauge coupling constant, 
this should be useful as an alternative 
approximation method instead of the expansion in 
inverse powers of gauge coupling. 
Our iterative approximation is at least guaranteed to be 
valid for asymptotic region $y \rightarrow \infty$, 
although it may wildly oscillate and possibly diverge 
near the wall. 
On the other hand, we have seen that the expansion in 
power of inverse gauge coupling does not converge 
near the wall nor asymptotically. 
To obtain a full solution, one needs to determine the 
integration constant such as $c_1$ which can be done 
by smoothly connecting to solutions from the region 
near the wall. 
Only when this integration constant is chosen to be a 
particular value, the solution should smoothly 
connect to the solution near the wall.

\renewcommand{\thesubsection}{Acknowledgements}
\subsection{}

One of the authors (N.S.) acknowledges 
the hospitality of the International Centre for 
Theoretical Physics at the last stage of this work. 
This work is supported in part by Grant-in-Aid 
 for Scientific Research from the Japan Ministry 
 of Education, Science and Culture 13640269 and 01350.
The work of K.O. is supported in part by Japan Society for
the Promotion of Science under the Post-doctoral Research Program.

\newcommand{\J}[4]{{\sl #1} {\bf #2} (#3) #4}
\newcommand{\andJ}[3]{{\bf #1} (#2) #3}
\newcommand{\AP}{Ann.\ Phys.\ (N.Y.)}
\newcommand{\MPL}{Mod.\ Phys.\ Lett.}
\newcommand{\NP}{Nucl.\ Phys.}
\newcommand{\PL}{Phys.\ Lett.}
\newcommand{\PR}{ Phys.\ Rev.}
\newcommand{\PRL}{Phys.\ Rev.\ Lett.}
\newcommand{\PTP}{Prog.\ Theor.\ Phys.}
\newcommand{\hep}[1]{{\tt hep-th/{#1}}}

\end{document}